\documentclass{article}
\usepackage{arxiv}
\usepackage[utf8]{inputenc} 
\usepackage[T1]{fontenc}    
\usepackage{hyperref}       
\usepackage{url}            
\usepackage{booktabs}       
\usepackage{amsfonts}       
\usepackage{nicefrac}       
\usepackage{microtype}      
\usepackage{graphicx}
\usepackage[round, authoryear]{natbib}
\usepackage{amsmath}
\usepackage{amssymb}
\usepackage{tikzit}

\usepackage{xcolor}	
\usepackage{tocloft}
\usepackage{epsfig}
\usepackage{graphics}
\usepackage{color}
\usepackage{soul}
\usepackage{ulem}

\usepackage{amsthm}

\newcommand{\suppmat}{Supplementary Material\ }
\newcommand{\sofr}{scalar-on-function regression\ }

\newcommand{\mgcv}{\texttt{mgcv}}

\newcommand{\rulesep}{\unskip\ \vrule\ }

\title{Adding structure to generalized additive models, with applications in ecology}

\author{ \href{https://orcid.org/0000-0002-9640-6755}{\includegraphics[scale=0.06]{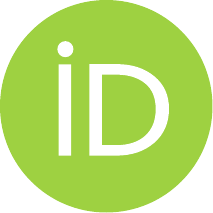}
\hspace{1mm}David L. Miller}\\
Biomathematics and Statistics Scotland\\
Dundee\\
Scotland\\
\\
UK Centre for Ecology \& Hydrology\\
Lancaster\\
United Kingdom\\
\texttt{dave.miller@bioss.ac.uk} \\
	\And
	\href{https://orcid.org/0000-0003-2738-4801}{\includegraphics[scale=0.06]{orcid.pdf}\hspace{1mm}Thomas Cornulier} \\
Biomathematics and Statistics Scotland\\
Aberdeen\\
Scotland\\	
	\texttt{thomas.cornulier@bioss.ac.uk} \\
 	\And
	\href{https://orcid.org/0000-0003-1734-5833}{\includegraphics[scale=0.06]{orcid.pdf}\hspace{1mm}Ken Newman} \\
Biomathematics and Statistics Scotland\\
Edinburgh\\
Scotland\\
\\
School of Mathematics\\
University of Edinburgh\\
Edinburgh\\
Scotland\\
	\texttt{ken.newman@bioss.ac.uk} \\
}

\renewcommand{\shorttitle}{Adding structure to GAMs}

\hypersetup{
pdftitle={Adding structure to generalized additive models, with applications in ecology},
pdfsubject={stat.ME},
pdfauthor={David L. Miller, Thomas Cornulier},
pdfkeywords={generalized additive modelling, linear functional, splines, smoothing},
}

\begin{document}
\maketitle

\begin{abstract}
Generalized additive models (GAMs) connecting a set of scalar covariates that map 1-1 to a response are commonly employed in ecology and beyond. However, covariates are often inherently non-scalar, taking multiple values for each observation of the response. They can sometimes have a temporal structure, e.g., a time series of temperatures, or a spatial structure, e.g., multiple soil pH measurements made at nearby locations. While aggregating or selectively summarizing such covariates to yield a scalar covariate allows the use of standard GAM fitting procedures, exactly how to do so can be problematic and information is necessarily lost. Naively including all $p$ components of a vector-valued covariate as $p$ separate covariates, say, without recognizing the structure, can lead to problems of multicollinearity, data sets that are excessively wide given the sample size, and difficulty extracting the primary signal provided by the covariate. Here we introduce three useful extensions to GAMs that efficiently and effectively handle vector-valued covariates without requiring one to choose aggregations or selective summarizations. These extensions are varying-coefficient, scalar-on-function and distributed lag models. While these models have existed for some time they remain relatively underused in ecology. This article aims to show when these models can be useful and how to fit them with the popular R package \mgcv{}.
\end{abstract}


\section{\label{Intro}Introduction}

Many statistical models that are currently widely-used (e.g., the linear model, generalized linear models, generalized additive models (GAMs) etc) make the assumption that explanatory data can only enter the model via scalar covariates (i.e., when one response observations maps to one observation of the covariate). Increasingly, we come into contact with more complex, structured predictor variables which require models that take into account that structure in order to adequately describe the phenomena in question. A commonly-encountered but simple structure is that of time series. It is often unclear what summary of these data is appropriate in a given situation. Researchers will often include one of: ($i$) the most proximate (in space/time) covariate value \citep[e.g.,][]{miller_estimating_2022}, ($ii$) an average based on information from previous studies \citep[e.g.,][]{frederiksenRegionalAnnualVariation2007} or ($iii$) multiple covariates at differing resolutions either at the same time or select one of these using some information criterion \citep[e.g.,][]{stuberBayesianMethodAssessing2017}. Ideally domain-specific knowledge guides this discussion, but it may be down to convenience, data availability or lack of information about the phenomenon in question. Na\"{i}ve alternatives to summarization, such as including all $p$ components of a vector-valued covariate as $p$ separate covariates without recognizing the structure, can lead to problems of multicollinearity, models with more parameters than data, and difficulty extracting the primary signal provided by the covariate. This article aims to inform readers about a suite of models that can help.

\begin{figure}
    \centering
    \includegraphics[width=0.32\textwidth]{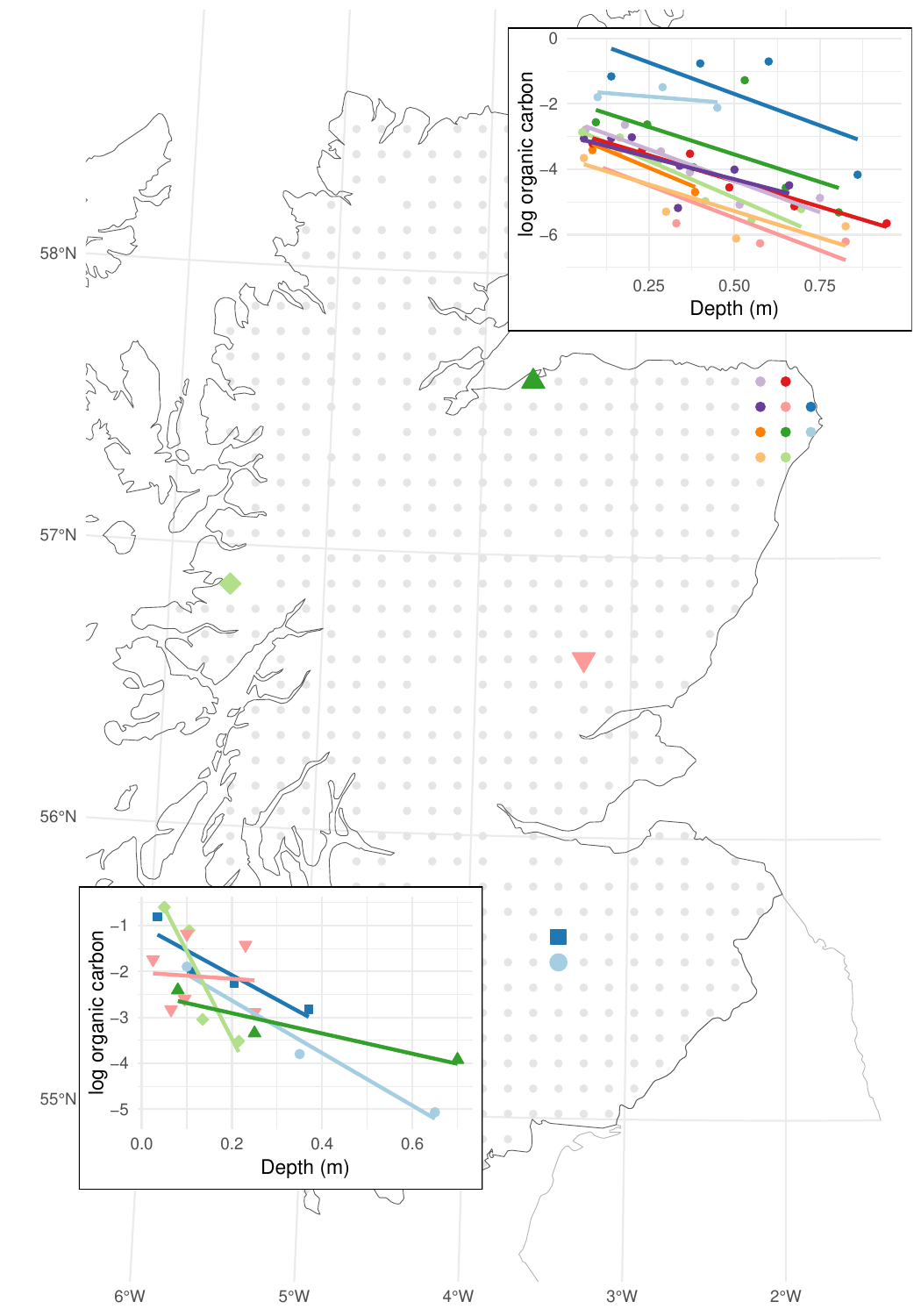}
    \rulesep
    \includegraphics[width=0.33\textwidth]{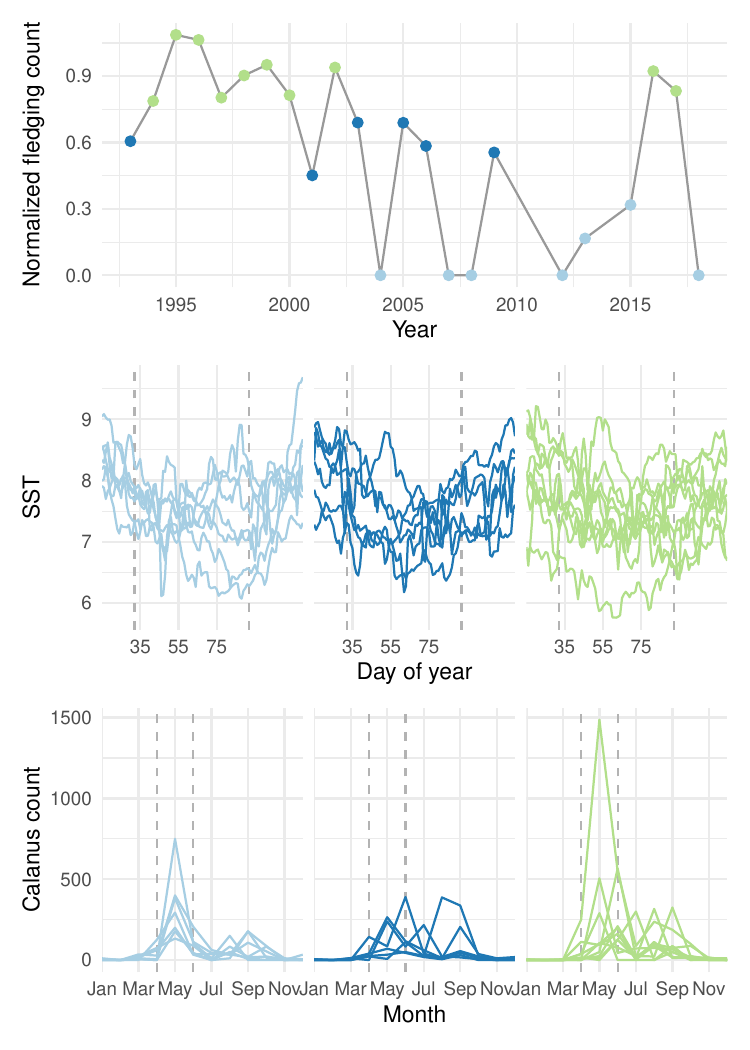}
    \rulesep
    \includegraphics[width=0.30\textwidth]{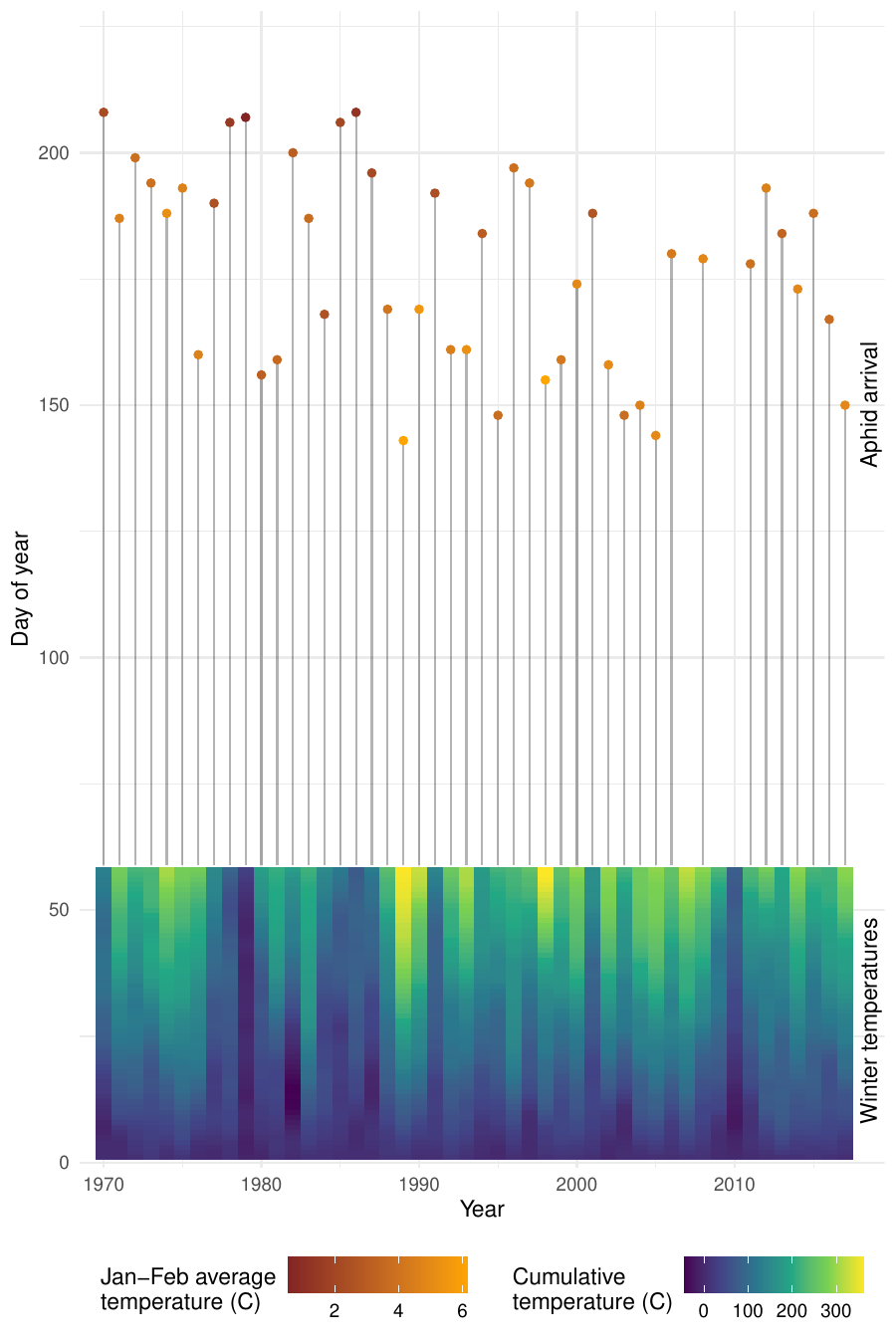}
    \caption{Three example situations where the models described in this paper can be applied. 
    Left: log carbon vs depth for several locations in the National Soil Inventory of Scotland; we expect a linear relationship with the slope varying smoothly in space (due to environment); samples close to one-another are more similar in their slope (top right subplot) than those that are randomly located (bottom left subplot). 
    Middle: top time series shows (normalized) yearly fledgling counts of kittiwakes at Gultak, Orkney; middle and lower plots show time series of sea surface temperatures (SSTs) each day of the year and of calanus counts per month, for multiple years; yearly fledgling counts are roughly classified as high (green), medium (dark blue) or low (light blue) with coloured dots in the top plot corresponding to the facets in the lower covariate plots; we know these are important drivers but it is not entirely obvious what is going on, so we build a model to gain some insight.
    Right: arrival of 10$^\text{th}$ aphid at the Dundee suction trap (points, coloured with average January/February temperature) with cumulative January-February temperature beneath; years that have higher cumulative temperatures (more green/yellow) tend to have earlier arrival dates (lower day of year value).}
    \label{fig:ex-cartoons}
\end{figure}

We consider three related models that allow us to bring more structure to our modelling. To introduce these ideas, here are three examples of the situations where these models would be useful:
\begin{itemize}
    \item Data from a systematic nation-wide survey of soil. At each geographical location we expect that the relationship between depth of sample and log carbon is linear (that carbon decreases in a log-linear manner with depth). We effectively want to fit a linear model per core.
    We wish to efficiently use parameters (i.e., regularize) and incorporate the spatial dependence in the data.
    The \textit{varying-coefficient model} allows the slope coefficient to vary according to another smooth (in this case, space).
      
    \item Data on breeding success for colonies of a seabird species that are central place foragers (i.e., a species which forages from a fixed location, e.g., a nest). Given year-to-year changes in climate and prey availability, it is not clear what if any single summarization of environmental covariate time series will adequately capture the relevant effects (e.g., using average spring prey availability is not always appropriate). Rather than aggregating at arbitrary scales, the \textit{scalar-on-function regression model} (also known as \textit{signal regression model}) estimates a smooth over the previous year's prey breeding season, giving weights to each month (via a smooth function). Now the product of the prey count and weight can be summed to determine the most explanatory aggregations during model fitting and calculate weighted sums of the covariates, based on the data.

    \item Modelling the emergence from diapause of aphid species which are agricultural pests. Temperature is thought to be an important mediator, particularly having several consecutive warm days coming out of winter. In this case we wish to model the cumulative effect of temperatures on date which aphids first take flight after diapause, however we do not know how many days of warm temperatures matter (nor do we know the relative effects of temperatures above a threshold). We could fit many models with different combinations of lagged temperature data, but this is complicated, time-consuming and subject to inconsistencies. The \textit{distributed lag model} allows us to simultaneously model data at all available lags and their interaction with time in a less subjective, more data-driven way.

\end{itemize}
Figure \ref{fig:ex-cartoons} shows brief slices of the of the situations described above, fully discussed throughout the paper.

In the statistical literature, multiple-valued covariates (that come from a single process) classically falls under \textit{functional data analysis} \citep{ramsay2013functional} which concerns itself with the case where the ``data'' are functions rather than single numbers.
Functional data analysis is a large area of statistics, encompassing a wide variety of techniques. In this paper, we are interested in those approaches that fall into the generalized additive modelling framework. 

Generalized Additive Models (GAMs) are an extension to the generalized linear model which allow for flexible semi-parametric modelling of covariates via structured random effects, splines, Gaussian Markov random fields, kriging-type approaches and more \citep{wood_generalized_2017-1}. We can write a basic GAM as
\begin{equation}
\mathbb{E}(y_i) = g^{-1}(\mathbf{w}_i\boldsymbol{\theta}   + \sum_j s_j(x_{ij})), ~~i=1,\ldots,n; ~ j=1,\ldots,J
\label{gambasic}
\end{equation}
where $y_i$ is the response which we model (on the scale of the link function, $g$) as an additive combination of: $w_i$, a vector of fixed effects covariates with corresponding parameters $\boldsymbol{\theta}$ to be estimated and $s_j$ are $J$ smooth terms of the covariates which can each be decomposed into a sum of $k$ basis functions multiplied by parameters: $s_j(x) = \sum_k \beta_k b_k(x)$. The response can be distributed according to some extended exponential family distribution. To ensure that we do not overfit with the very flexible smooth terms, we apply a penalty of the integrated squared derivatives of the $s_j$s to the likelihood to ensure that terms are ``just wiggly enough'' (equivalent to a multivariate normal prior on $\boldsymbol{\beta}$ with a given covariance structure; see \citet{millerBayesianViewsGeneralized} for an overview). For each $s_j$, the penalty (which is scaled by an estimated smoothing parameter) will reduce the degrees of freedom given to the smooth terms ($k$) to an effective degrees of freedom which is appropriate for the data; we therefore only need to set $k$ to be ``big enough'' and the penalization will estimate an appropriate degree of wigglyness for the term; \cite{pya_note_2016}.

GAMs are extremely flexible and have been applied in many areas of ecology (see references in \citet{millerBayesianViewsGeneralized}).
Here we showcase extension to \eqref{gambasic} which enable the modelling of the phenomena shown in Figure \ref{fig:ex-cartoons}.

The rest of the paper is structured as follows: in Section 2 we will show the basic theory behind extending \eqref{gambasic}. Section \ref{sec:vcm} describes the varying-coefficient model, which we can use to estimate a slope parameter that varies according to a function of another covariate, e.g., spatial location, and gives an application using data from the National Soil Inventory of Scotland. Section \ref{sec:sofr} describes the scalar-on-function regression model, which allows us to think of covariates as a functions of other covariates, which we can estimate the appropriate aggregations of during model fitting; we then illustrate the use of this class of models by applying it to estimate the breeding success of kittiwake colonies in Orkney, when we are not sure of the temporal scales which are important for model covariates. Section \ref{sec:dist-lag} introduces the distributed lag model, which allows us to jointly model the effects of environmental covariates and an index variable, e.g., temperature time series, at different lags and applies this model to aphid arrival dates in the UK. Finally, Section \ref{sec:disc} draws links between the models and gives directions for possible future research and application of these methods.


\section{Extending the generalized additive model}
\label{sec:extending}

We extend the generalized additive model by changing both the flexibility and the interpretation of the smooth terms, $s_j$, in the model. By changing how these terms contribute to the linear predictor we can invoke the three model types we will discuss here. 

To construct these terms, we introduce three bits of GAM machinery to help us. These are: ($i$) tensor products, ($ii$) the summation convention and ($iii$) ``by'' variables. These techniques are at the core of the approaches we will talk about later in the paper, so we give an overview here.

\subsection{Tensor products}
\label{sec:tensors}

We denote a univariate smooth of a variable $x$ as $s_x(x)$ and may extend this to two variables as $s_{xy}(x,y)$ (and so on) to create interaction terms between $x$ and $y$. But how do we construct $s_{xy}(x,y)$?

The most common way to construct the bivariate term $s_{xy}(x,y)$ is to use a multivariate basis like thin plate regression splines \citep{wood_thin_2003}. This basis has the (sometimes) unappealing property of \textit{isotropy} (or rotational invariance), meaning that all dimensions are treated as being on the same scale. This is a problem when we have $x$ be temperature in Celsius and $y$ be precipitation in millimetres. We need a smoother that will respect these differing scales.

The tensor product is a simple and general construction that takes multiple univariate smooths and uses them as the marginal terms in a multivariate smooth. This construction allows for smooth interactions of any number of covariates. The mathematical derivation of this is given in Appendix \ref{sec:app:tensor-pen}. In an \texttt{mgcv} model formula, we can include a tensor product by using \texttt{te()} in place of \texttt{s()} when constructing a multivariate term. 

To simplify notation, we do disambiguate smoother construction in mathematical notation (e.g., a 2-dimensional smooth may be denoted as simply $s(x,y)$, tensor or not), but do describe the construction of each term after the equation.

\subsection{The summation convention}
\label{sec:sum-conv}

Our second tool is a convention used in the R package \mgcv{}. If the arguments of an \texttt{s()} or \texttt{te()} term are matrices, the ``summation convention'' is invoked. This means that the resulting smooth term is summed over the columns of the matrices. For example, for a tensor product smooth of two matrices:
$$
s_{xy}(X_i, Y_i) = \sum_{q=1}^Q s_{xy}(X_{iq}, Y_{iq}),
$$
where $X_i$ indicates the $i^\text{th}$ sample (row of $X$) and $X_{iq}$ indicates the $i,q^\text{th}$ element of $X$, where the $Q$ columns contain multiple values of the covariate in question. The resulting values are then summed over the columns ($q$). Matrices $X$ and $Y$ must be of the same size.

The summation convention allows us to condense our time series covariates into a single value for use in the model's linear predictor.
\subsection{Linear functionals}

We can extend \eqref{gambasic} further by pre-multiplying the smooth terms by another object, $L_{ij}$ (the ``by'' variable, for observation $i$ and smooth term $j$). Different forms for $L_{ij}$ lead to different models.

In general we can re-write \eqref{gambasic} as:
\begin{equation}
\mathbb{E}(Y_i) = g^{-1}(\mathbf{w}_i \boldsymbol{\theta} + \sum_j L_{ij} s_j(x_{ij})).
\label{gamlf}
\end{equation}
The simplest form of the above is setting $L_{ij}$ to be an \textit{evaluation functional} (where $L_{ij} s_j(x_{ij})=s_j(x_{ij})$), which recovers \eqref{gambasic}. However, depending on whether $L_{ij}$ is a binary, continuous or factor scalar, or a vector, we obtain radically different models:
\begin{description}
    \item[Binary] by multiplying the smoother by either 0 or 1 we have the ability to include/remove the smooth for subsets of the data or predictions. This is useful from a practical perspective when we have an obvious temporal effect we wish to ``turn on'' for some time period (e.g., anthropogenic disturbance; \cite{jacobson_quantifying_2022}). 
    \item[Factor] Duplicates the smooth for each factor level, generating a type of \textit{factor-smooth interaction} model (``hierarchical GAM''; \citet{pedersen_hierarchical_2019}).
    \item[Numeric] For single numerical covariates, we just have $L_{ij}s_j(x_i) = z_{ij} s_j(x_i)$: a \textit{varying-coefficient model} \citep{trevor_hastie_varying-coefficient_1993}. We can think of $s$ as a model coefficient which is itself a function of another covariate (so we could write $z_{ij} \beta_j(x_i)$; hence varying-coefficient).
    \item[Vector] When $L_{ij}$ is a vector, we can think of $s_j(x_{ij})$ as a ``weight'' for each of the covariate values in $L_{ij}$, so $\sum_i L_{ij} s_j(x_{ij})$ is a weighted sum of the covariates, where the weights are smooth over the data index ($x_{ij}$) and are estimated during model fitting. We can view that summation as an approximation to the integral required for performing \textit{\sofr}. Integrating takes us from the many values of our functional data to a single summary measure over the full set of function values. 
\end{description}

In \texttt{mgcv} the \texttt{by=} argument to an \texttt{s()} or \texttt{te()}/\texttt{ti()} term gives the column or object we wish to use as our linear functional e.g., \texttt{s(x, by=y)}. The value of \texttt{y} can then be one of the four types above. For the vector case we can use a matrix-column in a \texttt{data.frame}. See \suppmat for more practical examples of setting-up these models (including how to set-up matrix-columns in \texttt{data.frame}s).

In the next three sections we investigate varying-coefficient models, \sofr and distributed lag models in detail, showing examples of their use and how to implement these models in R, using the three ``tricks'' we have just described.

\section{Varying-coefficient models (VCMs)}
\label{sec:vcm}

The varying-coefficient model \citep{trevor_hastie_varying-coefficient_1993} allows a slope coefficient to smoothly vary according to another variable. For example, letting a slope coefficient to vary with geographic location. To do this, we move from the case where the parameter is fixed (i.e., $\theta x$) to where it is a smooth of another covariate, $z$, giving $\theta(z)x$. We write the term as $\theta(z)x$ to be clear that this is a varying-coefficient, though we could reasonably also write it as $s(z) x$. These models not only add structure to the model, but also allows for inference about $\theta(z)$ directly, meaning we can look at e.g., how slopes vary in time or space. \cite{klappsteinStepSelectionFunctions2024} and \cite{thorsonSpatiallyVaryingCoefficients2023} give recent applications in ecology.


It is generally the case in practice that not only the slope, but also the intercept varies with the index. In which case it is usually necessary to include some effect of $z$ (a smooth of $z$) to ensure there is not misspecification. There may be cases where this is unnecessary, for example when random effects operating at a level lower than $z$ are included, so that the model is already flexible enough to capture intercept variation. The utility of adding a smooth of the index, $s(z)$ say, can be tested by simply looking at the effective degrees of freedom for that term and assessing whether $s(z)$ is flat.

The model makes the assumption that the effect varies smoothly in the index, though as we will see below, this still leads to some very flexible models. In addition, the smoothness of the function will tend to penalize excessive variation of the effects along the index, and thereby, will tend to stabilize the coefficients. In our soil carbon-depth situation, if we included a covariate for each (unevenly sampled) depth we would expect correlations between those coefficients, which might destabilize the fitting. The smooth function therefore plays two important roles, which are to support richer inference and to make model estimates more reliable, provided that the smoothness assumption is valid.

In estimating $\theta(z)$ we are estimating a continuous surface of slopes for the covariate $x$. So, looking at this surface gives us an idea where trends are positive or negative. This is quite different from the normal interpretation of a smoother in a model (which represents the whole effect size of the term), but leads to an extremely useful summary of (effectively) many related regressions.






\subsection{Setting-up varying-coefficient models in \mgcv{}}

Setting-up a VCM in \texttt{mgcv} requires we have our index covariate, \texttt{z}, and the (numeric) linear functional, \texttt{x}, as columns in our data.  We can then write our term as \texttt{s(z, by=x)} in our model formula. As shown in the following example we can specify the rest of the smooth via arguments to \texttt{s()} as usual.


\subsection{Example: modelling the National Soil Inventory of Scotland using varying-coefficient models}

The organic carbon stored in soil (soil organic carbon; SOC) depends on the land use: e.g., crop residues in agricultural fields and leaf litter in forests. Changing from one land use to another (e.g., creating more agricultural land by cutting down a forest for fields, or ``rewilding'' existing farmland by planting trees) leads to changes in carbon storage in soil. To measure the effect of these land use changes, we need model soil organic carbon as a function of both location and depth.

\begin{figure}
    \centering
    \includegraphics[width=\linewidth]{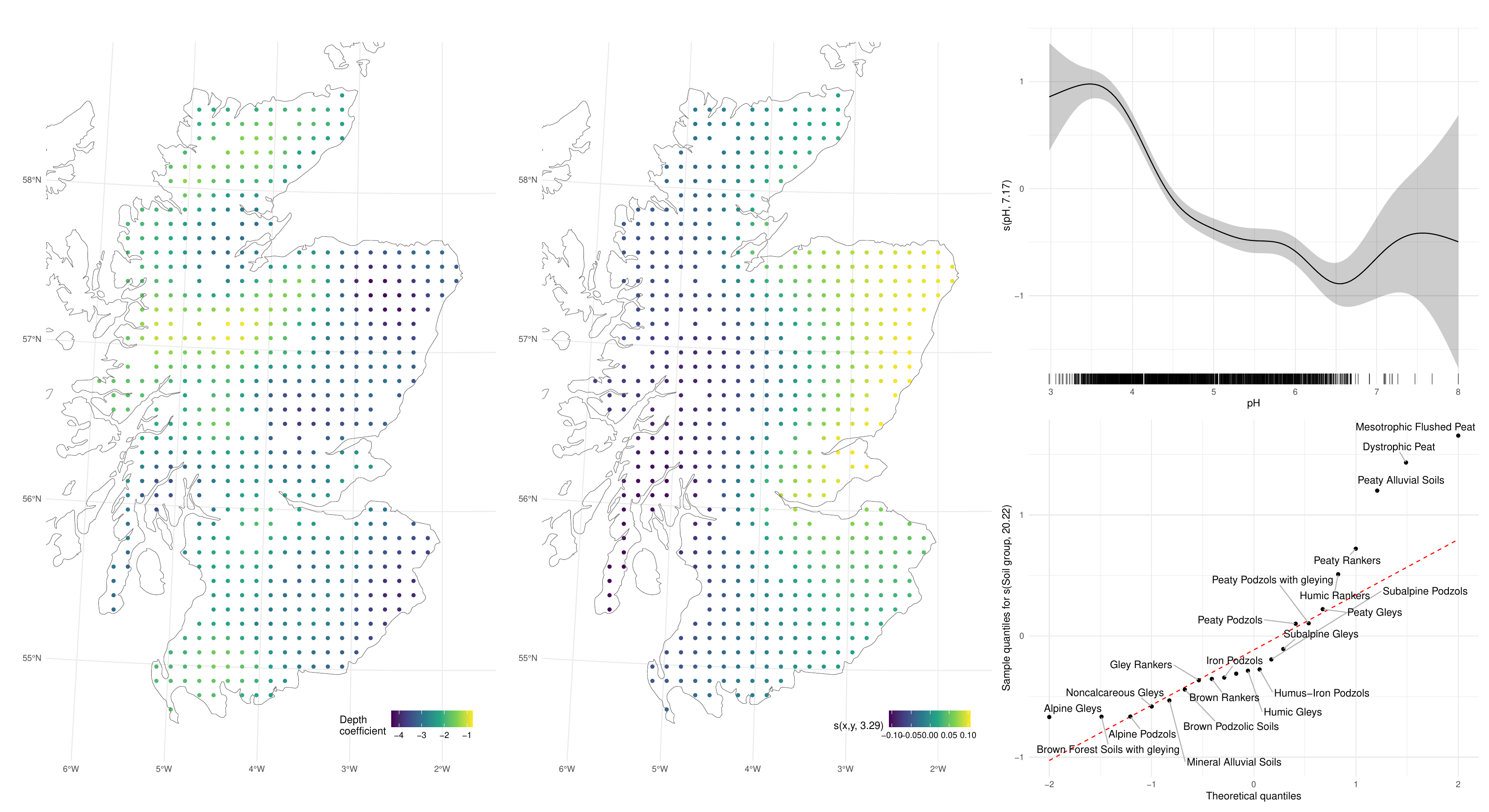}
    \caption{Term plots for our soil organic carbon model. Left: coefficient estimates, $s_z(x,y)$), for the varying-coefficient term in our model of soil organic carbon. Coefficient values give the slope of the linear relationship between $\log_e$ carbon concentration and depth. Middle: spatial smoother, $s_{xy}(x,y)$ giving the ``baseline'' carbon. Right, top: smooth of pH, showing decreasing effect on carbon as soil becomes less acidic. Right, bottom: quantile-quantile plot of the random effect of soil subtype, showing relatively good agreement between the model assumptions of normality and estimates. The unlabelled point in the middle of the final plot is "Brown Forest Soils".}
    \label{fig:vc_coefs_plot}
\end{figure}

The National Soil Inventory of Scotland \cite[NSIS;][]{lillyNationalSoilInventory2010} data were collected between 1978 and 1988 on a 10km grid across Scotland, where sampling locations where at the intersections (Figure \ref{fig:vc_coefs_plot}). At each point a soil pit was dug, the main horizons (layers) were identified and sampled. The samples were generally taken from a 10cm band around the mid-depth of each horizon where possible and avoiding contamination of the horizons above or below. There are 721 sample locations in Scotland. For the analysis presented here, we only use those samples on the Scottish mainland (609 locations).

To measure the amount of organic carbon that is captured in soil, we can take soil samples at different depths (the number of depths used varies in number and magnitude but is usually 3--10, up to a depth of about 1m). In the case of NSIS, carbon was measured using a Hewlett-Packard CHN 185 analyser (Hewlett-Packard, Palo Alto, CA, United States) where the soil sample is combusted at high temperature in the presence of oxygen. Carbon dioxide produced during combustion of carbon is determined using a thermal conductivity detector. We measure SOC as carbon concentration in units of mass carbon per unit of mass ``bulk'' soil (so, for example, grams per kilogram).

We are interested in the relationship between carbon concentration and depth. \cite{jobbagyVerticalDistributionSoil2000} show that $\log_e$ carbon \textit{density} will decrease linearly with depth and Figure \ref{fig:ex-cartoons} (left panel) shows that this seems to hold for carbon concentration too. We can then build a varying-coefficient model where the coefficient for depth varies in space. We also need to account for other important environmental variables: soil type and pH (at depth). We also include a spatial smooth to account for any general change in SOC in space. There has been some work on spatial modelling of SOC in the soil science literature, for example \cite{poggio_national_2014} build a 3D model of SOC using a combination of GAMs and kriging.

We formulate the following model:
\begin{equation}
    \log c_i = \beta_0 + s_z(x_i,y_i)z_i  + s_{x,y}(x_i,y_i) + \beta_{S,i} + s_P(P_i) + \epsilon_i, \quad\text{for i=1,\ldots, 1922}
\end{equation}
where $c_i$ is the carbon concentration in sample $i$, which is a unique combination of location and depth, $z_i$ (to a total of 1922 samples). $s_z(x_i,y_i)$ is the corresponding spatially-varying coefficient, $s_{x,y}(x_i,y_i)$ is a spatial smooth (where $x_i,y_i$ gives the projected coordinates of observation $i$), $\beta_{S,i}$ is a normally-distributed, mean zero random effect on the soil type (which does not vary by depth), $s_P(P_i)$ is a smooth of pH (at depth $z_i$) and $\epsilon_i$ are independent, identically distributed normal errors. $\beta_0$ is an intercept. We model $\log_e c_i$ as normally distributed. Since we are smoothing over an area with a boundary (the coastline of mainland Scotland), we model $s_{x,y}(x_i,y_i)$ and $s_z(x_i,y_i)$ using the soap film smoother \citep{wood_soap_2008} to ensure that there is no ``leakage'' between unrelated areas.


A full analysis is given in \suppmat A, including example code to run the model, diagnostics and other useful information. Figure \ref{fig:vc_coefs_plot} shows the resulting per-term plots for our model. Of primary interest to us is the left plot of the varying-coefficient term. Values give the slope of the linear effect of depth. We see that the coefficients with smaller absolute value (yellow), which indicate more carbon storage at larger depths, are concentrated in the Highland area. Coastal areas are more moderate and the North East (Aberdeenshire), where the absolute value of the coefficient is very large (dark blue), have very steep carbon drop-off with depth. The model enforces our assumption that nearby sample locations have similar depth relationships, such that this varies smoothly in space. The rest of Figure \ref{fig:vc_coefs_plot} shows the other terms in the model: the general trend in carbon concentration over mainland Scotland (middle plot), the effect of soil pH and a quantile-quantile plot for the random effects, labelled for the different soil subtypes. Note that the outliers in the quantile-quantile plot are predominantly peats (which store more water).

This relatively simple model allows us to include two important pieces of prior knowledge: that samples that are close together are generally more similar (spatial parts of the model) and the assumption of a linear relationship between $\log_e$ carbon concentration and depth. 

\section{Scalar-on-function regression (SoFR)}
\label{sec:sofr}

If we want to model, say, the breeding success of a seabird as a function of environmental covariates (Figure \ref{fig:ex-cartoons}), in conventional linear/additive models we are constrained to the use of a single covariate for each of those environmental variables we deem useful. \cite{frederiksenRegionalAnnualVariation2007} used sea surface temperature (SST) averaged over February and March in the previous year to each survey to summarize the contribution of sea surface temperature to breeding success. We might think of other summaries we can make, perhaps including both February and March means separately. Taking this to the limit we might then include the full year time series of daily SST in the model. Our model would then include the sequence of terms $ \beta_{\texttt{SST}, 1}\texttt{SST}_{1} + \ldots + \beta_{\texttt{SST}, 1}\texttt{SST}_{365})$. Clearly this model has far too many (correlated) coefficients and would likely suffer from multicolinearity issues. Information criterion-based term selection in such a situation would be arduous at best.

Scalar-on-function regression (signal regression) \citep{goldsmithPenalizedFunctionalRegression2011, mcleanFunctionalGeneralizedAdditive2014, reissMethodsScalaronFunctionRegression2017}, allows us to fit a model where we use a full series of covariates per sample. We use a parallel smooth over the covariate index to ``weight'' the covariate series, before summing over the time period in question. This creates summaries that are based on the data, rather than ones that are fixed prior to modelling.

The central conceptual move in SoFR is to consider the covariate as a \textit{function}, rather than discrete observations. For example, rather than thinking of SST as multiple discrete values, corresponding to different times, (e.g., $\texttt{SST}_{t}$), we instead think of it as a function that we are evaluating at a given time: $\texttt{SST}(t)$. This conceptual switch lets us then estimate a corresponding smooth over $t$, $s(t)$ which tells us the importance of the times of the SST measurements.

Generally let the variable of interest $x$ (e.g., SST) varying according to some index variable $v$ (e.g., day of year), so we write $x(v)$. We estimate a smooth $s(v)$, a function of an index variable $v$. The SoFR model is as follows:
\begin{equation}
\int_\mathcal{D} x_i(v) s(v) \text{d}v,
\label{eq:sofr}
\end{equation}
where for each observation $i$ we have a complete realisation of the functional $x_i(v)$ (there is one complete function per observation; e.g., the time series of temperature for a given point in space). $s(v)$ is sometimes referred to as the \textit{probe} or \textit{contrast template}, which gives the weighting for every point in $\mathcal{D}$ (domain over which we are interested in $v$, e.g., the time interval, spatial region, etc). Rather than asking ``how does covariate $x$ affect the response?'', we are asking ``how much does the covariate affect the response at $v$, when taken as part of the average over $\mathcal{D}$?''.

To calculate the SoFR term in practice, we approximate \eqref{eq:sofr} by quadrature:
\begin{equation}
\int_\mathcal{D} x_i(v) s(v) \text{d}v \approx \sum_j x_i(v_j) s(v_j),
\label{eq:sofr-approx}
\end{equation}
where $j$ indexes the quadrature points: a grid over $\mathcal{D}$ where $x$ is available. Relating the right-hand side of (\ref{eq:sofr}) to the framework described in Section \ref{sec:extending}, we can see that the pre-multiplication of the smooth by the covariate is handled by a vector linear functional. Then since we can ``stack'' the vectors to make a matrix, we will then invoke the summation convention and obtain the form given here. This allows us to estimate the SoFR term using regular GAM methods.

Figure \ref{fig:kitti-terms} shows example $x_i(v)$ and $s(v)$ for the kittiwake data we will analyse in the final part of this section. In the case of both sea surface temperature (SST) and calanus (a small crustacean, which is prey to sandeels, the prey of kittiwakes) count, $v$ is time (though on daily and monthly scales, respectively). In each case the data ($x(v)$) are initially very messy, the smooths ($s(v)$) are simpler, picking out time periods that are relevant, leading to $x(v)s(v)$ that is more interpretable (see descriptions in the analysis).




\subsection{Setting-up scalar-on-function regression models in \mgcv{}}

To generate \eqref{eq:sofr} we need to invoke both a vector linear functional and the summation convention: we will use the \texttt{by} argument and matrix covariates. We can fit $\sum x(v) s(v)$ by writing \texttt{s(V, by=X)} if \texttt{V} is the (matrix) covariate that indexes the (matrix) covariate (\texttt{X}). 
\subsection{Modelling kittiwake breeding success using scalar-on-function regression}

Black-legged kittiwakes (\textit{Rissa tridactyla}; henceforth, kittiwakes) are  seabirds that nest on cliffs to breed. During this time, kittiwakes feed on small forage fish (e.g., \textit{Ammodytes tobianus}, the lesser sandeel) which constitute a key part of the diet of chicks, so the availability and quality of sandeels is expected to correlate with annual breeding success \citep{dauntForagingStrategiesBlacklegged2002}. We analyse data on the number of successful fledglings raised by kittiwakes per year in Orkney (an archipelago 80km off the North coast of Scotland) collected as part of the Seabird Monitoring Programme (\url{https://www.bto.org/our-science/projects/seabird-monitoring-programme}).

Unfortunately data on sandeel abundance are typically limited in their spatial and temporal resolution; for example e.g., \cite{langtonVerifiedDistributionModel2021} provide a spatial but not temporal model. Sandeel abundance is thought to be mediated by two proxy covariates: sea surface temperature (SST) as a proxy for environmental conditions and calanus (small crustacean) counts as a proxy for prey \citep{frederiksenRegionalAnnualVariation2007}. We obtained SST data from the National Oceanic and Atmospheric Administration's web services (specifically product \texttt{ncdcOisst2Agg\_LonPM180}; \cite{huangImprovementsDailyOptimum2021}). Calanus data were obtained from the Marine Biological Association as part of their Continuous Plankton Recorder (CPR) dataset \citep{ostleCPRDataOSPAR2021}, these are raw counts from surveys so, following \cite{frederiksenRegionalAnnualVariation2007} we took monthly averages. Spatially, we took data from within 80km of the polygons defining the islands of Orkney, suggested as a maximum foraging range for kittiwakes \citep{dauntForagingStrategiesBlacklegged2002}. See \suppmat B for data preparation.

\cite{frederiksenRegionalAnnualVariation2007} used SST averaged over February/March of the previous year, whereas for calanus April-June of the current year or January to August of the previous year were used. Literature suggests that temporal overlap in sandeel and calanus development is important in sandeel recruitment \citep{regnierUnderstandingTemperatureEffects2019}, so the time period used to calculate the summary statistic may well be sensitive to climate effects on a yearly basis. SoFR gives a flexible approach to this problem where we don't know which months are important in a given year. We would expect a sum of calanus and potentially sea surface temperature would be influential in the sandeel abundance and hence kittiwake breeding success (see Discussion for more on choosing the right structure). SoFR allows us to include the full time series of the data, weighted by the smoothers to build a data-based picture of the influence of these effects.

Our model for kittiwake takes the following form:
\begin{equation}
\mathbb{E}(n_{k,y}) = \exp \left[ \log(A_{k,y}) + \sum_{m=1}^{18} \texttt{calanus}_{m,y} s(m)+ \sum_{d=15}^{120} \texttt{SST}_{d,y-1} s(d) + s_y(y) + \beta_k\right],
\label{eq:kitti}
\end{equation}
where $n_{k,y}$ is the number of fledglings at site $k$ in year $y$ and $A_{k,Y}$ is the corresponding number Apparently Occupied Nests (AON) in that year/site combination (effectively a measure of effort). We assume that the fledgling count, $n_{k,y}$, is negative binomial distributed; we estimate a scale parameter, $\kappa$ (controlling the mean-variance relationship $\text{Var}(X) = \mathbb{E}(X) + \kappa \mathbb{E}(X)^2$). Calanus counts from 
the 18 months from May in year $y$ to January in year $y-1$, the effect is moderated by a smooth of month $m$. The sea surface temperature from the previous year $\texttt{SST}_{d,y-1}$ is moderated by a smooth of day $d$. $s_y(y)$ is a smooth of year and $\beta_k$ is a normally-distributed random intercept for site. The use of 18 months of calanus counts in our model allows us to effectively perform model selection during fitting. The temporal smooth of year captures the trend that is common between all colonies. The full analysis is reported in \suppmat B.

\begin{figure}[t]
    \centering
    \includegraphics[width=0.95\linewidth]{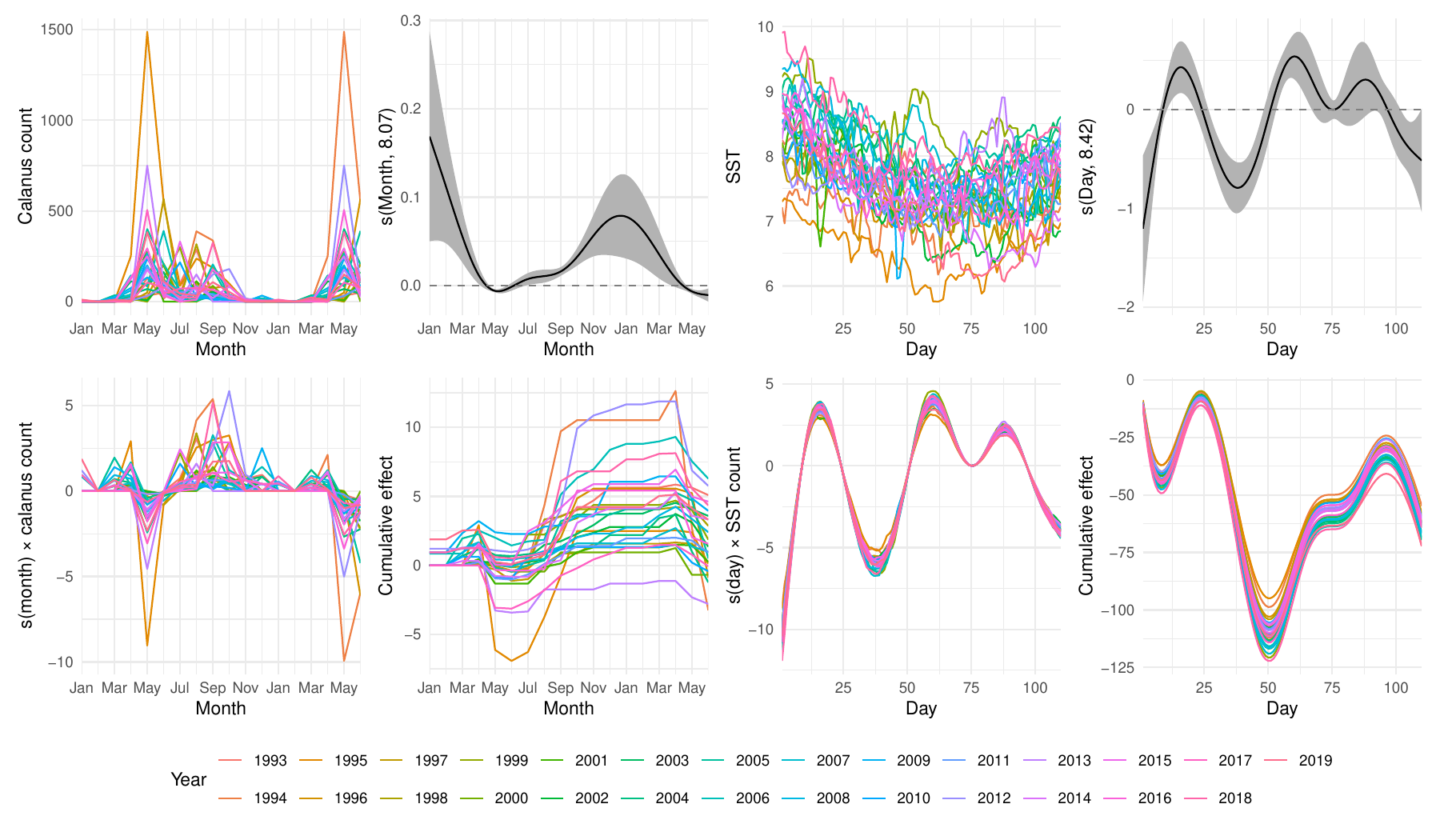}
    \caption{Data and resulting smooths from our analysis of kittiwake breeding success. The left four plots relate to the calanus term the right four plots to sea surface temperature. In each case, from top left subplot, clockwise: the data, the smooth of the index variable, the product of data and the smooth, the cumulative sum of the product of the data and smooth. The far right value of the latter plot being the value that enters the linear predictor.}
    \label{fig:kitti-terms}
\end{figure}

Figure \ref{fig:kitti-terms} shows various plots which we think illustrate parts of the model. The right four plots relate to the calanus effect, the left four to SST. In both cases, we see the raw data in the top left subplot, right of that is the estimated smooth. The bottom row shows the product of the smooth and the data followed by the cumulative effect (i.e., the cumulative sum of the product). Only the furthest right value of the cumulative plot enters the model, however we find that plot useful to have an idea of how the effect changes over time. In both cases we can see that by the time we get to the cumulative plot, the effects are much smoother and more regular than the original data.

The left four plots are related to the calanus term. The data show increasing calanus counts from January, peaking in May/June with positive counts through to November in most years. Our smoother shows the most uncertainty in the times where calanus counts are very low (comparing left and right plots). Combining the data and smooth (bottom left) we see that we have large negative contributions corresponding to the times where calanus is highest but positive contributions while calanus is positive after that main peak. The cumulative plot (bottom right) shows fairly flat start, with some years having a decrease May through July, but after that an increasing effect through to October when things level out (no additional effects, corresponding to times when calanus counts are very low). Finally we see a reduction in the effect at the end of the series, corresponding to the negative effects of the large calanus counts in the current year (echoing the \citep{frederiksenRegionalAnnualVariation2007} negative effect for same-year calanus).

The right four plots are SST-related. The smoother plot shows sign change over the series, giving a variable effect over the days included in the analysis. Although this covariate may be harder to interpret, we do see a large change in AIC by adding this term to the model ($\Delta$AIC=163).

We compared model (\ref{eq:kitti}) with the averaged covariates as described in \cite{frederiksenRegionalAnnualVariation2007} (also including a smooth of year and site random effect, as above) and found that the corresponding average models had lower AICs ($\Delta$AIC $>$20 between our \sofr and best average model) and deviances explained ($>$ 35\% difference in both cases). The average-based models also did not find that calanus to be a significant covariate to include in the model (approximate $p$-value of 0.15 for previous year calanus, 0.42 for current year). 

\begin{figure}[t]
    \centering
    \includegraphics[width=0.8\linewidth]{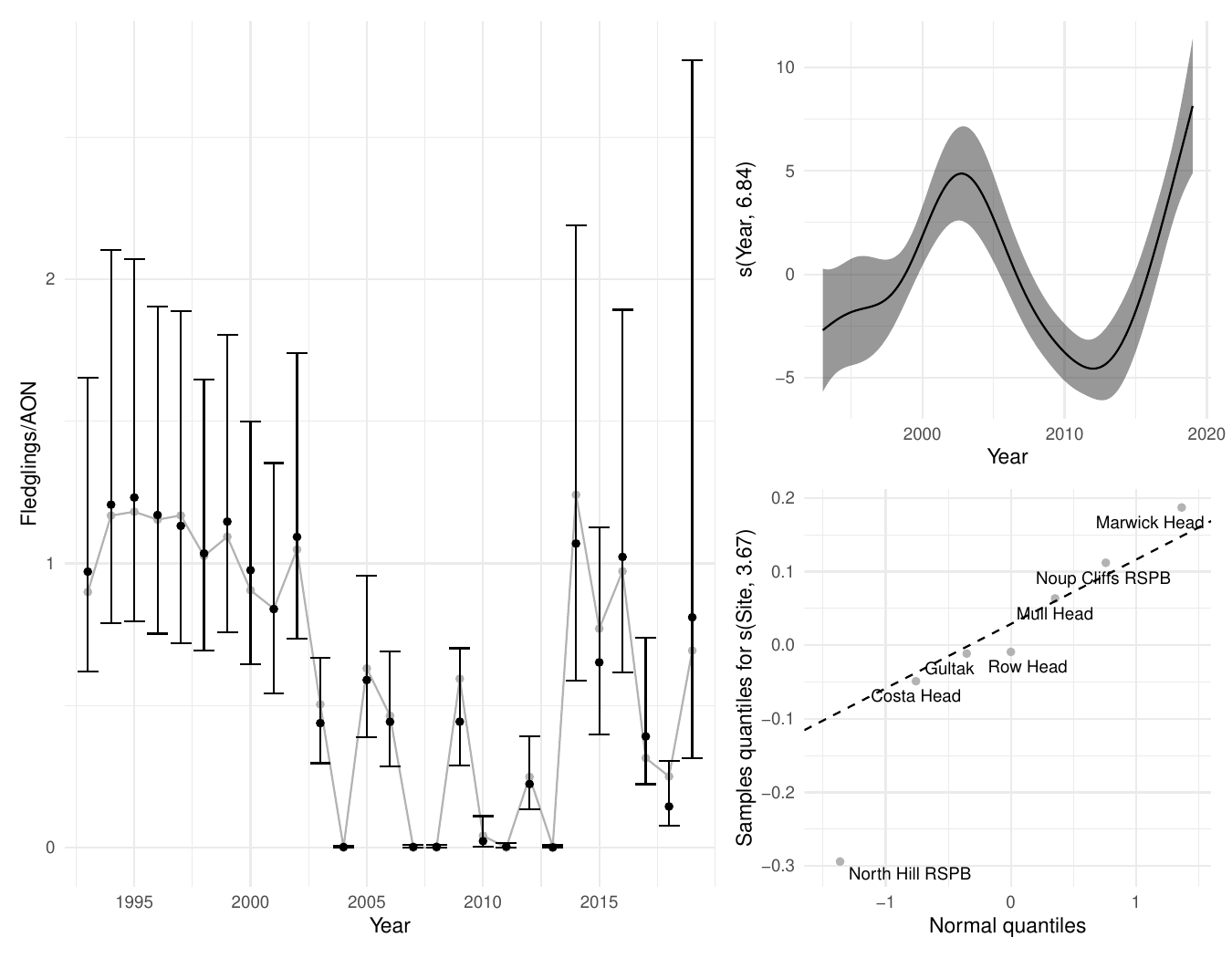}
    \caption{Results from \sofr modelling of kittiwakes. Left: kittiwake breeding success trend; grey points show the counts summed over the sites, joined by the grey line for clarity; black dots and whiskers show the model estimates and 95\% credible intervals, generated from posterior sampling of our model. Values were normalized by dividing by the apparently occupied nests (AON). Right top: the year term from our model, which captures the general trend in breeding success. Bottom right: quantile-quantile plot for the site random effect in our model; we can see that North Hill RSPB is a considerable outlier.}
    \label{fig:all-ork-trend}
\end{figure}

Our model describes all Orcadian sites so we compare the all-Orkney trend from the model to the data (with uncertainty calculated using posterior sampling per \citet{millerBayesianViewsGeneralized}). Figure \ref{fig:all-ork-trend}) (left) shows good agreement between the model and data. The temporal smooth (right top plot) shows the general trend over all sites in time. Given the small area considered by the model, this seems valid and illustrates the dip with trend in kittiwake success in the mid 2000s to mid 2010s. Finally we show the quantile-quantile plot for the random effect on site. This effect was relatively small but we can see that the most northerly site, North Hill RSPB, is an outlier compared to the other sites. \suppmat B provides per-site trends and comparison with other models.

Our use of scalar-on-function regression models here has allowed us to include the effect of SST and calanus count time series to describe the breeding success of kittiwakes in Orkney. \suppmat B describes how to setup the data for such a model.

\section{Distributed lag models (DLMs)}
\label{sec:dist-lag}

The SoFR model assumes that the effects are just the weighted sum of the covariates, rather than looking at the interaction between the index variable and the covariate of interest. Distributed lag models \citep{armstrongModelsRelationshipAmbient2006, gasparriniDistributedLagLinear2011} build on the ideas of SoFR by creating a two dimensional smooth of the covariate of interest and its index (usually referred to as a \textit{lag}, as it classically refers to time). This allows us to model e.g., the effect of temperatures and date \textit{jointly} on aphid emergence (Figure \ref{fig:ex-cartoons}), not just weighting which times are relevant.

To construct our interaction we use a tensor product (since we will almost always have different scales for the covariate and index) and then the summation convention to accumulate the effect. We can write the model as $\sum_{t=1}^T s_{xt}(x_{t}, t)$ where $s(\cdot)$ is a tensor product of the covariate and lag. We have $T$ (equally-spaced) lags of our covariate $x_{t}$ to include in our model. The lag values, $t$, are lags up to the maximum $T$. A detailed derivation is given in Appendix \ref{dist-lag-derivation}.

\subsection{Setting-up distributed lag models in \mgcv{}}

We can write our model term as \texttt{te(X, LAG)} where \texttt{X} and \texttt{LAG} are matrices with as many columns as lags ($T$) that are of interest and each row giving one observation's data (so $n$ rows). Note that it is important to make this term a tensor product as the variables \texttt{X} and \texttt{LAG} are measured on different scales (and hence are not isotropic).

The invocation of the summation convention is automatic when the covariates are matrices, so model construction from the \texttt{mgcv} perspective is easy. The more difficult part in practice is the sourcing, processing, and arrangement of the matrices.

\subsection{Modelling aphid arrival using distributed lags}

Many species of aphids (superfamily \textit{Aphidoidea}) are agricultural pests that are
vectors for the transmission of plant viruses that cause devastating 
and consequential economic damage to hundreds of varieties of
crops on a global scale \citep{harringtonEnvironmentalChangePhenology2007}.  Various mitigation strategies are used in agriculture around the world (including insecticides, introduction of natural predators, etc), though since their use tends to be time-sensitive, accurate forecasts of aphid ``arrival" in fields are necessary to be most effective.

Rothamsted Insect Survey (RIS) and Science and Advice for Scottish Agrigulture (SASA; part of the Scottish Government) jointly operate a network of suction traps located across the United Kingdom that capture insects in flight, including aphids, on a near daily basis. This network of (currently) 16 suction traps 
(\url{https://insectsurvey.com/about})
provides information that can be used to model aphid emergence from their overwintering state (either as insects or eggs, thanks to the aphids' ability to reproduce both parthenogenetically or sexually).

One of the most popular models that predict the timing of the earliest catches of aphids is a simple linear regression using average temperature during January and February \citep{turlApproachForecastingIncidence1980}. These models are used by RIS and SASA to inform farmers and agronomists (e.g., \url{https://insectsurvey.com/aphid-bulletin}).


In light of the developmental dynamics of insects as a function of temperatures, the use of January-February average temperatures is a quite coarse summarization, although such summaries can at times still yield useful and relatively accurate predictions (enough to be used in practice; e.g., \cite{harringtonNewApproachUse1991}). Distributed lag models allow for a more complete use of daily temperature data with potentially improved predictions as well as insight into notions underlying development rates.

To demonstrate the distributed lag model we look at just one species of aphids: \textit{Sitobion avenae}, the English grain aphid. It is thought to mostly use an anholocylic reproductive strategy \citep{waltersOverwinteringStrategyTiming1986}, so most reproduction is asexual, e.g., females produce clones of themselves. Overwintering occurs as young or adults (in grasses) and are therefore metabolically sensitive to temperature. 

We consider the distributed lag model extension of the classic January-February average temperature linear regression:
\begin{equation}
\label{eq:Dist.Lag.Aphid.model}
\mathbb{E}(A_{iu}) = \beta_0 + \sum_{t=1}^{59} s(\tau_{iut},t),
\end{equation}
where $A_{iu}$ is the day-of-year (where 1 corresponds to January 1) when the 10$^\text{th}$ aphid was caught  in year $i=1970, \ldots, 2017$ at the suction trap in Dundee, Edinburgh or Newcastle ($u=1,2,3$). We assume $A_{iu}$ is normally distributed. The term $ s(\tau_{iut}, t)$ is a tensor product smooth of daily temperatures at each year, site and lag ($\tau_{iut}$) and the lags themselves ($t$). Dundee, Edinburgh and Newcastle are situated on the (drier, sunnier) East coast of the UK. Newcastle is located in the North East of England, whereas Edinburgh is in the ``Central Belt'' of Scotland $\sim$80km North of Newcastle, and Dundee $\sim$60km North of Edinburgh.

\begin{figure}[t]
    \centering
    \includegraphics[width=1\linewidth]{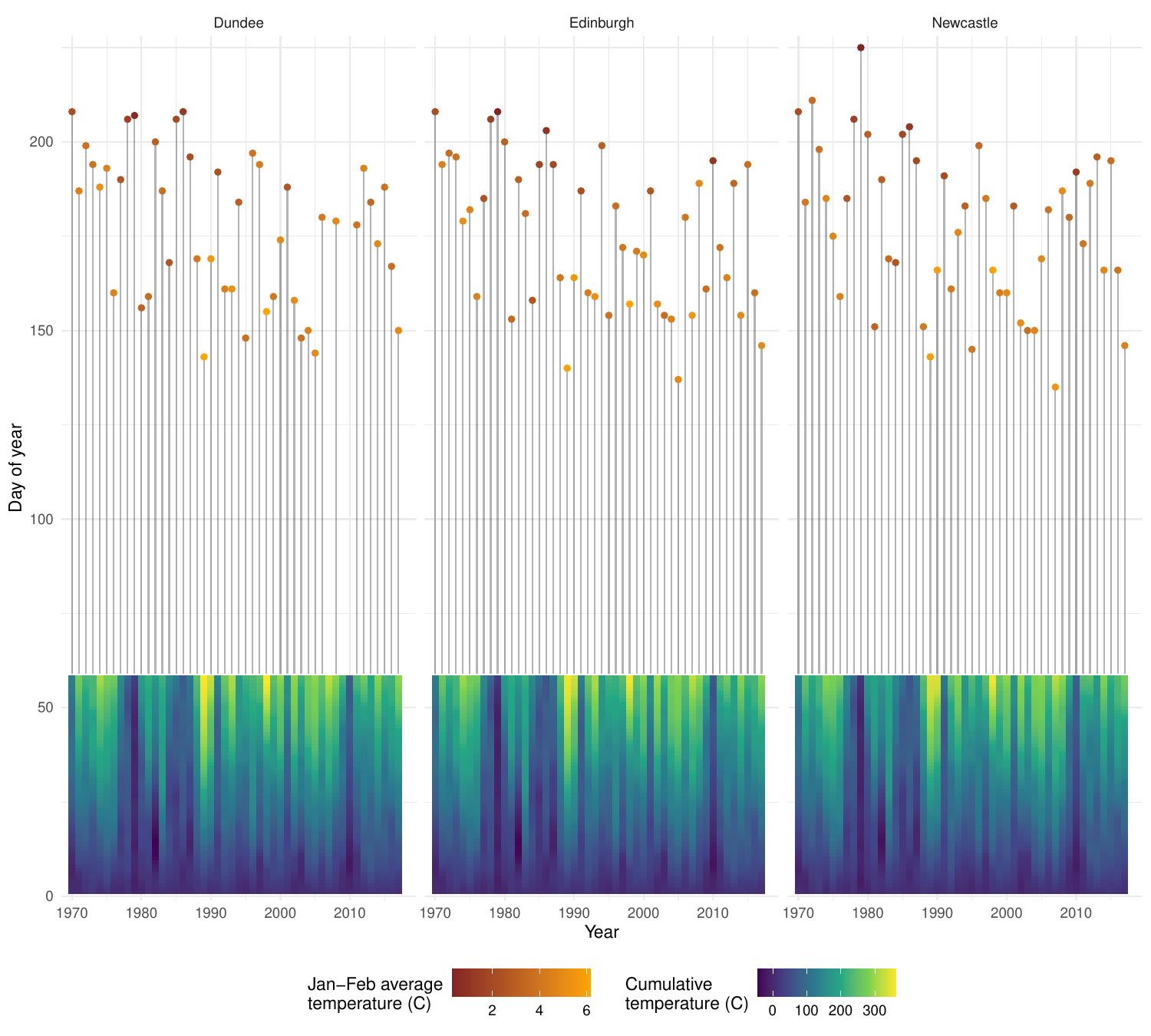}
    \caption{Temperature time series and arrival dates for the $10^\text{th}$ catch of \textit{Sitobion avenae} at the suction traps in Dundee, Edinburgh and Newcastle. The vertical axis shows day of the year; the bottom segment shows the cumulative average daily temperature from January $1\text{st}$ through February $28^\text{th}$, followed by the arrival date given by the dots (lines to aid registration between temperatures and arrivals), dot colours give the average January-February temperatures. We generally see that earlier arrivals (lower positions of dots on the vertical axis) correlate with milder average temperatures (yellower dots), however note that there is more complexity in the cumulative time series below. Our distributed lag model aims to account for this by directly modelling the time series.}
    \label{fig:savenae_arrival_temps}
\end{figure}

We obtained daily average temperatures from January 1 through to February 28 for the three sites by extracting data from the Met Office's HadUK-Grid product \citep{hollisHadUKGridANewUK2019}. This provides an interpolation between the Met Office's weather station network. We show the arrivals, temperature time series and the average January-February temperatures in Figure \ref{fig:savenae_arrival_temps}.

\begin{figure}[t]
  \centering
  \includegraphics[width=1\linewidth]{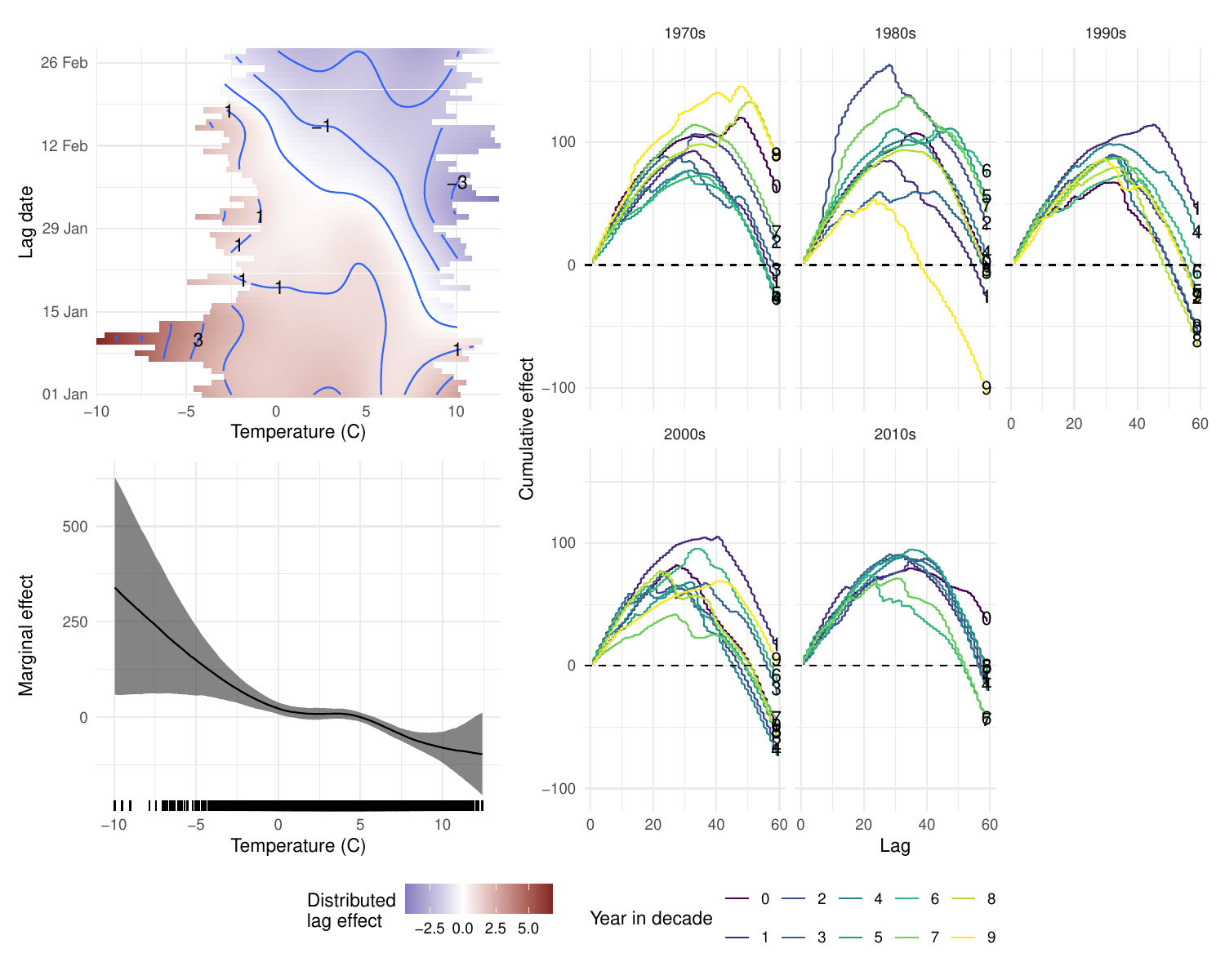}
  \caption{Three plots showing different views of the distributed lag effect of temperature and lag for \textit{Sitobion avenae}. The top left plot shows the $s(\tau, t)$ effect in two dimensions over the range of temperatures observed at each lag. The bottom left plot shows $\sum_{t=1}^{59} s(\tau,t)$, the temperature effect, marginalised over time lags, with uncertainty (grey band). The right of the figure shows the cumulative effects over lags, faceted by decade for clarity; the far right value of each line indicates the effect size that enters the linear predictor. Note that since we use an identity link in our model, effect sizes are on the response scale, so we can read them in terms of days (positive values indicate delay, negative indicates advancement of arrival day).}
  \label{F:Temperature.Lag.aphid.arrival}
\end{figure}

The smooth of daily January and February temperatures and lag is plotted in Figure \ref{F:Temperature.Lag.aphid.arrival} in three different ways to show the different features of the method. In all cases, effect sizes are on the response scale, so we can read them on the day scale, negative values indicating that arrival was earlier than the intercept value (175.8, around 25$^\text{th}$ June), positive values meaning arrival was later.

The top left plot of \ref{F:Temperature.Lag.aphid.arrival} shows the tensor $s(\tau, t)$ over the range of temperatures observed at each lag (over all sites). The main feature of this plot is the almost diagonal zero contour indicating that the effect of temperature is positive (i.e., delay) at low temperatures, even relatively late in the time series. This effect diminishes as temperature increases, leading through to negative effects (i.e., earlier arrival) if we have temperatures $\approx10^\circ$C around halfway through January. In the bottom left of the plot we see a large effect from very cold temperatures, which we might expect as \textit{Sitobion avenae}'s overwintering strategy means that they are sensitive to low temperatures, as they are not protected by an egg during diapause.

The bottom left plot of Figure \ref{F:Temperature.Lag.aphid.arrival} the temperature effect marginalised over time lags (the sum the ``columns" in the tensor plot). The mean effect shows  cold temperatures cause later arrivals, warmer temperatures moderate this effect, before giving a moderate change from the mean. Clearly colder temperatures have more drastic effects than warmer ones. This may at least in part be down to the additional effect of increasing photoperiod as the year advances, leading to budding in their host plants and their migration to cereal crops. The uncertainty (calculated using posterior sampling) on this effect is particularly large at very cold temperatures, at least partially due to the small number of observations at these low temperatures (rug plot). Indeed, experiments suggest that the median lethal temperature is around $-8^\circ$C for \textit{Sitobion avenae} \citep[][Table 2.2]{williamsOverwinteringLowTemperature1983}.

The cumulative effects over lags are shown in the right side of Figure \ref{F:Temperature.Lag.aphid.arrival}. The plot is faceted by decade and labelled year-within-decade for each line. These plots allow us to see how similar or different each year was in terms of its effect trajectory, considering temperatures only implicitly. For example, we can see that the 1970s and 1980s have especially wide ranges of final effect sizes, whereas the 2000s and especially the 2010s appear more concentrated. The winter of 1977-1978/1978-1979 in the UK was known as the ``winter of discontent'' because of various sociopolitical issues exacerbated by severe storms and the coldest winter in 16 years (two highest yellow lines in the 1970s plot). On the other hand, 1988-1989 was the warmest on record at the time (lowest and most dramatic yellow line in the 1980s plot). 

\suppmat C gives further analysis, model checking and comparison with the January-February average model currently in use and another model that uses weekly averages.

Overall we find that the distributed lag approach gives significant, interpretable insight into aphid arrivals. The development of these additional plots can help the practitioner understand what is happening in their data more effectively by investigating the results through different facets.


\section{Discussion}
\label{sec:disc}

We have illustrated three related extensions of the GAM methodology which can add beneficial structure to our models. Structure is the important feature here: modellers must think carefully about the mechanisms that drive the response in their systems. This (hopefully) moves us away from the ``everything, including the kitchen sink'' ``framework'' that has been popular in applied statistical ecology, where model selection (be that via information criteria or otherwise) is left to clean-up the mess. We hope that these models will prompt some introspection and careful thought about modelling, especially given the relative speciality and complexity of the terms.

Adding complexity and structure to our models generally entails a considerable burden when it comes to number of parameters used and therefore the amount of data needed to fit them. The datasets used here are relatively modest in size (soil carbon data, 1922 observations at 609 locations; aphid data 141 observations at three traps, kittiwake data 136 observations). In each case the linear functional term used a modest number of parameters. These are reasonable when compared to the parameters required for fitting models to spatial or temporal subsets of the data, or by aggregating covariates to finer groupings.

The use of index variables allows us to use covariates on differing scales within one model. This avoids us either having to average or use another model to interpolate to get everything on the same scale. Indeed, as we showed for the kittiwake analysis, we can include two variables (SST and calanus count) with two differently-scaled indices (day and month) with no issue. The summation over the indices mitigates a mild-to-serious headache for the modeller.

Taking naive summaries potentially disrupts the smooth process: we expect day-to-day changes in temperature to be small, but taking monthly or seasonal averages breaks this correlation somewhat. When we ``chunk'' the covariate in this way, we may also limit the number of unique values (limiting available degrees of freedom).

\subsection{Making links}

Coming from a the standard GAM formula in \eqref{gambasic}, the varying-coefficient model is an extension of the random slope model to where there is structure in the covariance matrix for the random effect---in this case the structure is that of a spline.

We can see the \sofr model as an extension to the varying-coefficient model, as we have the same underlying terms (values of the covariate multiplied by a smoother) but in the signal regression case we average multiple values for each sample over the range of covariate values. In the varying-coefficient case we just have one value of the covariate per sample.

The distributed lag and \sofr models are very similar, both rely on the summation convention to construct their terms. In the distributed lag case the summation happens over the tensor product, so the smoother is applied first, whereas in the \sofr case we are using the ``raw'' values of the covariate. In both cases we can get a lot of information out of looking at the cumulative effects (summing over the index) as we do in Figures \ref{fig:kitti-terms} and \ref{F:Temperature.Lag.aphid.arrival}.

Similarities between the models we have discussed are highlighted in Table \ref{tab:model-comparison}, where we show the mathematical definition for each model, along with R code and some detail on the structure of the variables used in the model.

\begin{table}
    \centering
\begin{tabular}{l|lll}
    & Definition & \texttt{formula} component  & Variable structure\\
\hline 
GAM & $s(x_i)$  & 
 \texttt{s(x)} & \texttt{x} scalar\\
VCR   & $\theta_{z}(z_i)x_i$  &
 \texttt{s(z,by=x)} & \texttt{x}, \texttt{z} scalar, \texttt{z} index\\
SoFR  & $ \sum_z s(z) x_{i,z}$ &
 \texttt{s(Z,by=X)} & \texttt{X}, \texttt{Z} are matrices\\
   &  & & \texttt{X}, \texttt{Z} matrix of stacked $x$, $z$ \\
 DLM  & $\sum_{z} s(x_{i,z},z) $ & \texttt{te(X, Z)} &
  \texttt{X}, \texttt{Z} are matrices\\
   &  & & \texttt{X}, \texttt{Z} matrix of stacked $x$, $z$ \\
\end{tabular}
    \caption{Comparison of the models reviewed in this paper, along with the ``basic'' GAM structure. Their simplest formulations in mathematics and R code are shown along with a reminder on the shape of the variables that are used. 
    }
    \label{tab:model-comparison}
\end{table}


\subsection{Further extensions}

The example analyses given here show simple situations where these approaches can be used, but they are only the beginning of the kinds of models that can be developed.

We can employ smoother bases with useful properties to better model the data structure. Just as we need to respect the coastline of Scotland if we smooth over the land, if a varying-coefficient model's index is actually based on some periodic phenomenon, then we can use a cyclic spline to model that. The \texttt{mgcv} package has many options for this and we suggest that users investigate the \texttt{mgcv::smooths} manual page for more information.

We can extend the distributed lag and SoFR models to spatial and temporal indices. Techniques from \cite{pedersen_hierarchical_2019} can be used to create hierarchical distributed lag models, sharing of information between sites, while still allowing for per-site differences. This would allow greater flexibility and the potential to address more complex ecological questions.

We can use multiple terms in one model. For example, we might improve our soil carbon model by adding a SoFR term which allows us to look at appropriate means of rainfall that take into account local conditions at each site. We are also not limited to including these terms in a GAM (or GAMM), since they have direct translations into structured Bayesian priors \citep{millerBayesianViewsGeneralized}, they can be included in the linear predictors of more complex models via probabilistic programming languages. For example we might want to embed the distributed lag formulation in a Bayesian population dynamics model for aphids.


Overall these models are just parts of the overall GAM and broader modelling landscape that can be used to improve our models of ecological phenomena. Having these additional tools, which enable us to include additional structure, gives us more flexibility to not only model phenomena more accurately but also make richer inferences.

\section*{Acknowledgements}

This paper is part of an ongoing framework agreement between BioSS and UKCEH. We thank all those involved in these projects for their input into the modelling process and those who attended workshops on this material for their input and feedback. In particular: Kate Searle, Maria Bogdanova, Charlotte Regan, Pete Levy, Fiona Seaton and Pete Henrys at UKCEH. The authors also wish to thank Allan Lilly (James Hutton Institute) for his explanation of the NSIS data. Katharine Preedy and John Addy (BioSS) provided useful suggestions and encouragement on the aphid example. Ali Karley (James Hutton Institute), Fiona Highet and Mairi Carnegie (SASA) gave vital background on the suction trap network and aphid biology. John Addy and Jackie Potts provided very helpful comments on an early version of the paper.

Funding for KN was provided by the Scottish Government's Rural and Environment Science and Analytical Services (RESAS) Division Strategic Research Programme 2022-2027 "Biodiversity and Ecosystem Tools" (BioSS-1-BET).

\bibliography{dave} 

\begin{thebibliography}{34}
\providecommand{\natexlab}[1]{#1}
\providecommand{\url}[1]{\texttt{#1}}
\expandafter\ifx\csname urlstyle\endcsname\relax
  \providecommand{\doi}[1]{doi: #1}\else
  \providecommand{\doi}{doi: \begingroup \urlstyle{rm}\Url}\fi

\bibitem[Armstrong(2006)]{armstrongModelsRelationshipAmbient2006}
B.~Armstrong.
\newblock Models for the {{Relationship Between Ambient Temperature}} and
  {{Daily Mortality}}.
\newblock \emph{Epidemiology}, 17\penalty0 (6):\penalty0 624--631, Nov. 2006.
\newblock ISSN 1044-3983.
\newblock \doi{10.1097/01.ede.0000239732.50999.8f}.

\bibitem[Daunt et~al.(2002)Daunt, Benvenuti, Harris, Dall, Elston, and
  Wanless]{dauntForagingStrategiesBlacklegged2002}
F.~Daunt, S.~Benvenuti, M.~P. Harris, A.~L. Dall, D.~A. Elston, and S.~Wanless.
\newblock Foraging strategies of the black-legged kittiwake {{Rissa}}
  tridactyla at a {{North Sea}} colony: Evidence for a maximum foraging range.
\newblock \emph{Marine Ecology Progress Series}, 245:\penalty0 239--247, Dec.
  2002.
\newblock ISSN 0171-8630, 1616-1599.
\newblock \doi{10.3354/meps245239}.

\bibitem[Frederiksen et~al.(2007)Frederiksen, Edwards, Mavor, and
  Wanless]{frederiksenRegionalAnnualVariation2007}
M.~Frederiksen, M.~Edwards, R.~A. Mavor, and S.~Wanless.
\newblock Regional and annual variation in black-legged kittiwake breeding
  productivity is related to sea surface temperature.
\newblock \emph{Marine Ecology Progress Series}, 350:\penalty0 137--143, Nov.
  2007.
\newblock ISSN 0171-8630, 1616-1599.
\newblock \doi{10.3354/meps07126}.

\bibitem[Gasparrini(2011)]{gasparriniDistributedLagLinear2011}
A.~Gasparrini.
\newblock Distributed {{Lag Linear}} and {{Non-Linear Models}} in
  {{{\emph{R}}}} : {{The Package}} {\textbf{dlnm}}.
\newblock \emph{Journal of Statistical Software}, 43\penalty0 (8), 2011.
\newblock ISSN 1548-7660.
\newblock \doi{10.18637/jss.v043.i08}.

\bibitem[Goldsmith et~al.(2011)Goldsmith, Bobb, Crainiceanu, Caffo, and
  Reich]{goldsmithPenalizedFunctionalRegression2011}
J.~Goldsmith, J.~Bobb, C.~M. Crainiceanu, B.~Caffo, and D.~Reich.
\newblock Penalized {{Functional Regression}}.
\newblock \emph{Journal of Computational and Graphical Statistics}, 20\penalty0
  (4):\penalty0 830--851, Jan. 2011.
\newblock ISSN 1061-8600, 1537-2715.
\newblock \doi{10.1198/jcgs.2010.10007}.

\bibitem[Harrington et~al.(1991)Harrington, Howling, Bale, and
  Clark]{harringtonNewApproachUse1991}
R.~Harrington, G.~G. Howling, J.~S. Bale, and S.~Clark.
\newblock A new approach to the use of meteorological and suction trap data in
  predicting aphid problems.
\newblock \emph{EPPO Bulletin}, 21\penalty0 (3):\penalty0 499--505, Sept. 1991.
\newblock ISSN 0250-8052, 1365-2338.
\newblock \doi{10.1111/j.1365-2338.1991.tb01281.x}.

\bibitem[Harrington et~al.(2007)Harrington, Clark, Welham, Verrier, Denholm,
  Hull{\'e}, Maurice, Rounsevell, Cocu, and
  Consortium]{harringtonEnvironmentalChangePhenology2007}
R.~Harrington, S.~J. Clark, S.~J. Welham, P.~J. Verrier, C.~H. Denholm,
  M.~Hull{\'e}, D.~Maurice, M.~D. Rounsevell, N.~Cocu, and E.~U.~E. Consortium.
\newblock Environmental change and the phenology of {{European}} aphids.
\newblock \emph{Global Change Biology}, 13\penalty0 (8):\penalty0 1550--1564,
  2007.
\newblock ISSN 1365-2486.
\newblock \doi{10.1111/j.1365-2486.2007.01394.x}.

\bibitem[Hastie and Tibshirani(1993)]{trevor_hastie_varying-coefficient_1993}
T.~Hastie and R.~Tibshirani.
\newblock Varying-{{Coefficient Models}}.
\newblock \emph{Journal of the Royal Statistical Society. Series B
  (Methodological)}, 55\penalty0 (4):\penalty0 757--796, 1993.

\bibitem[Hollis et~al.(2019)Hollis, McCarthy, Kendon, Legg, and
  Simpson]{hollisHadUKGridANewUK2019}
D.~Hollis, M.~McCarthy, M.~Kendon, T.~Legg, and I.~Simpson.
\newblock {{HadUK-Grid}}---{{A}} new {{UK}} dataset of gridded climate
  observations.
\newblock \emph{Geoscience Data Journal}, 6\penalty0 (2):\penalty0 151--159,
  2019.
\newblock ISSN 2049-6060.
\newblock \doi{10.1002/gdj3.78}.

\bibitem[Huang et~al.(2021)Huang, Liu, Banzon, Freeman, Graham, Hankins, Smith,
  and Zhang]{huangImprovementsDailyOptimum2021}
B.~Huang, C.~Liu, V.~Banzon, E.~Freeman, G.~Graham, B.~Hankins, T.~Smith, and
  H.-M. Zhang.
\newblock Improvements of the {{Daily Optimum Interpolation Sea Surface
  Temperature}} ({{DOISST}}) {{Version}} 2.1.
\newblock \emph{Journal of Climate}, 34\penalty0 (8):\penalty0 2923--2939, Apr.
  2021.
\newblock ISSN 0894-8755, 1520-0442.
\newblock \doi{10.1175/JCLI-D-20-0166.1}.

\bibitem[Jacobson et~al.(2022)Jacobson, Henderson, Miller, Oedekoven, Moretti,
  and Thomas]{jacobson_quantifying_2022}
E.~K. Jacobson, E.~E. Henderson, D.~L. Miller, C.~S. Oedekoven, D.~J. Moretti,
  and L.~Thomas.
\newblock Quantifying the response of {{Blainville}}'s beaked whales to
  {{U}}.{{S}}. naval sonar exercises in {{Hawaii}}.
\newblock \emph{Marine Mammal Science}, 38\penalty0 (4):\penalty0 1549--1565,
  Oct. 2022.
\newblock ISSN 0824-0469, 1748-7692.
\newblock \doi{10.1111/mms.12944}.

\bibitem[Jobb{\'a}gy and Jackson(2000)]{jobbagyVerticalDistributionSoil2000}
E.~G. Jobb{\'a}gy and R.~B. Jackson.
\newblock The {{Vertical Distribution}} of {{Soil Organic Carbon}} and {{Its
  Relation}} to {{Climate}} and {{Vegetation}}.
\newblock \emph{Ecological Applications}, 10\penalty0 (2):\penalty0 423--436,
  2000.
\newblock ISSN 1939-5582.
\newblock \doi{10.1890/1051-0761(2000)010[0423:TVDOSO]2.0.CO;2}.

\bibitem[Klappstein et~al.(2024)Klappstein, Michelot, Fieberg, Pedersen, and
  Mills~Flemming]{klappsteinStepSelectionFunctions2024}
N.~J. Klappstein, T.~Michelot, J.~Fieberg, E.~J. Pedersen, and
  J.~Mills~Flemming.
\newblock Step selection functions with non-linear and random effects.
\newblock \emph{Methods in Ecology and Evolution}, 15\penalty0 (8):\penalty0
  1332--1346, 2024.
\newblock ISSN 2041-210X.
\newblock \doi{10.1111/2041-210X.14367}.

\bibitem[Langton et~al.(2021)Langton, Boulcott, and
  Wright]{langtonVerifiedDistributionModel2021}
R.~Langton, P.~Boulcott, and P.~Wright.
\newblock A verified distribution model for the lesser sandeel {{Ammodytes}}
  marinus.
\newblock \emph{Marine Ecology Progress Series}, 667:\penalty0 145--159, June
  2021.
\newblock ISSN 0171-8630, 1616-1599.
\newblock \doi{10.3354/meps13693}.

\bibitem[Lilly et~al.(2010)Lilly, Bell, Hudson, Nolan, and
  W]{lillyNationalSoilInventory2010}
A.~Lilly, J.~Bell, G.~Hudson, A.~Nolan, and T.~W.
\newblock National {{Soil Inventory}} of {{Scotland}} ({{NSIS}} 1978-88), Mar.
  2010.

\bibitem[McLean et~al.(2014)McLean, Hooker, Staicu, Scheipl, and
  Ruppert]{mcleanFunctionalGeneralizedAdditive2014}
M.~W. McLean, G.~Hooker, A.-M. Staicu, F.~Scheipl, and D.~Ruppert.
\newblock Functional {{Generalized Additive Models}}.
\newblock \emph{Journal of Computational and Graphical Statistics}, 23\penalty0
  (1):\penalty0 249--269, Jan. 2014.
\newblock ISSN 1061-8600, 1537-2715.
\newblock \doi{10.1080/10618600.2012.729985}.

\bibitem[Miller(2025)]{millerBayesianViewsGeneralized}
D.~L. Miller.
\newblock Bayesian views of generalized additive modelling.
\newblock \emph{Methods in Ecology and Evolution}, n/a\penalty0 (n/a), 2025.
\newblock ISSN 2041-210X.
\newblock \doi{10.1111/2041-210X.14498}.

\bibitem[Miller et~al.(2022)Miller, Becker, Forney, Roberts, Ca{\~n}adas, and
  Schick]{miller_estimating_2022}
D.~L. Miller, E.~A. Becker, K.~A. Forney, J.~J. Roberts, A.~Ca{\~n}adas, and
  R.~S. Schick.
\newblock Estimating uncertainty in density surface models.
\newblock \emph{PeerJ}, 10:\penalty0 e13950, Aug. 2022.
\newblock ISSN 2167-8359.
\newblock \doi{10.7717/peerj.13950}.

\bibitem[Ostle(2021)]{ostleCPRDataOSPAR2021}
C.~Ostle.
\newblock {{CPR}} data for {{OSPAR PH1 FW5 PH2 PH3 FW2 QSR2023}}, 2021.

\bibitem[Pedersen et~al.(2019)Pedersen, Miller, Simpson, and
  Ross]{pedersen_hierarchical_2019}
E.~J. Pedersen, D.~L. Miller, G.~L. Simpson, and N.~Ross.
\newblock Hierarchical generalized additive models in ecology: An introduction
  with mgcv.
\newblock \emph{PeerJ}, 7:\penalty0 e6876, May 2019.
\newblock ISSN 2167-8359.
\newblock \doi{10.7717/peerj.6876}.

\bibitem[Peng et~al.(2009)Peng, Dominici, and
  Welty]{pengBayesianHierarchicalDistributed2009}
R.~D. Peng, F.~Dominici, and L.~J. Welty.
\newblock A {{Bayesian Hierarchical Distributed Lag Model}} for {{Estimating}}
  the {{Time Course}} of {{Risk}} of {{Hospitalization Associated}} with
  {{Particulate Matter Air Pollution}}.
\newblock \emph{Journal of the Royal Statistical Society Series C: Applied
  Statistics}, 58\penalty0 (1):\penalty0 3--24, Feb. 2009.
\newblock ISSN 0035-9254.
\newblock \doi{10.1111/j.1467-9876.2008.00640.x}.

\bibitem[Poggio and Gimona(2014)]{poggio_national_2014}
L.~Poggio and A.~Gimona.
\newblock National scale {{3D}} modelling of soil organic carbon stocks with
  uncertainty propagation --- {{An}} example from {{Scotland}}.
\newblock \emph{Geoderma}, 232--234:\penalty0 284--299, Nov. 2014.
\newblock ISSN 0016-7061.
\newblock \doi{10.1016/j.geoderma.2014.05.004}.

\bibitem[Pya and Wood(2016)]{pya_note_2016}
N.~Pya and S.~N. Wood.
\newblock A note on basis dimension selection in generalized additive
  modelling.
\newblock \emph{arXiv preprint arXiv:1602.06696}, 2016.

\bibitem[Ramsay and Silverman(2013)]{ramsay2013functional}
J.~Ramsay and B.~Silverman.
\newblock \emph{Functional Data Analysis}.
\newblock Springer Series in Statistics. Springer New York, 2013.
\newblock ISBN 978-1-4757-7107-7.

\bibitem[R{\'e}gnier et~al.(2019)R{\'e}gnier, Gibb, and
  Wright]{regnierUnderstandingTemperatureEffects2019}
T.~R{\'e}gnier, F.~M. Gibb, and P.~J. Wright.
\newblock Understanding temperature effects on recruitment in the context of
  trophic mismatch.
\newblock \emph{Scientific Reports}, 9\penalty0 (1):\penalty0 15179, Oct. 2019.
\newblock ISSN 2045-2322.
\newblock \doi{10.1038/s41598-019-51296-5}.

\bibitem[Reiss et~al.(2017)Reiss, Goldsmith, Shang, and
  Ogden]{reissMethodsScalaronFunctionRegression2017}
P.~T. Reiss, J.~Goldsmith, H.~L. Shang, and R.~T. Ogden.
\newblock Methods for {{Scalar-on-Function Regression}}: {{Scalar-on-Function
  Regression}}.
\newblock \emph{International Statistical Review}, 85\penalty0 (2):\penalty0
  228--249, Aug. 2017.
\newblock ISSN 03067734.
\newblock \doi{10.1111/insr.12163}.

\bibitem[Stuber et~al.(2017)Stuber, Gruber, and
  Fontaine]{stuberBayesianMethodAssessing2017}
E.~F. Stuber, L.~F. Gruber, and J.~J. Fontaine.
\newblock A {{Bayesian}} method for assessing multi-scale species-habitat
  relationships.
\newblock \emph{Landscape Ecology}, 32\penalty0 (12):\penalty0 2365--2381, Dec.
  2017.
\newblock ISSN 0921-2973, 1572-9761.
\newblock \doi{10.1007/s10980-017-0575-y}.

\bibitem[Thorson et~al.(2023)Thorson, Barnes, Friedman, Morano, and
  Siple]{thorsonSpatiallyVaryingCoefficients2023}
J.~T. Thorson, C.~L. Barnes, S.~T. Friedman, J.~L. Morano, and M.~C. Siple.
\newblock Spatially varying coefficients can improve parsimony and descriptive
  power for species distribution models.
\newblock \emph{Ecography}, 2023\penalty0 (5):\penalty0 e06510, May 2023.
\newblock ISSN 0906-7590, 1600-0587.
\newblock \doi{10.1111/ecog.06510}.

\bibitem[Turl(1980)]{turlApproachForecastingIncidence1980}
L.~A.~D. Turl.
\newblock An approach to forecasting the incidence of potato and cereal aphids
  in {{Scotland}}.
\newblock \emph{EPPO Bulletin}, 10\penalty0 (2):\penalty0 135--141, July 1980.
\newblock ISSN 0250-8052, 1365-2338.
\newblock \doi{10.1111/j.1365-2338.1980.tb02635.x}.

\bibitem[Walters and Dewar(1986)]{waltersOverwinteringStrategyTiming1986}
K.~F.~A. Walters and A.~M. Dewar.
\newblock Overwintering {{Strategy}} and the {{Timing}} of the {{Spring
  Migration}} of the {{Cereal Aphids Sitobion}} avenae and {{Sitobion}}
  fragariae.
\newblock \emph{Journal of Applied Ecology}, 23\penalty0 (3):\penalty0
  905--915, 1986.
\newblock ISSN 0021-8901.
\newblock \doi{10.2307/2403943}.

\bibitem[Williams(1983)]{williamsOverwinteringLowTemperature1983}
C.~T. Williams.
\newblock \emph{Overwintering and Low Temperature Biology of Cereal Aphids}.
\newblock PhD thesis, University of Southampton, 1983.

\bibitem[Wood(2003)]{wood_thin_2003}
S.~N. Wood.
\newblock Thin plate regression splines.
\newblock \emph{Journal of the Royal Statistical Society: Series B (Statistical
  Methodology)}, 65\penalty0 (1):\penalty0 95--114, 2003.

\bibitem[Wood(2017)]{wood_generalized_2017-1}
S.~N. Wood.
\newblock \emph{Generalized {{Additive Models}}. {{An Introduction}} with
  {{R}}}.
\newblock Texts in {{Statistical Science}}. CRC Press, 2nd edition, 2017.

\bibitem[Wood et~al.(2008)Wood, Bravington, and Hedley]{wood_soap_2008}
S.~N. Wood, M.~V. Bravington, and S.~L. Hedley.
\newblock Soap film smoothing.
\newblock \emph{Journal of the Royal Statistical Society: Series B (Statistical
  Methodology)}, 70\penalty0 (5):\penalty0 931--955, Nov. 2008.
\newblock ISSN 13697412, 14679868.
\newblock \doi{10.1111/j.1467-9868.2008.00665.x}.

\end{thebibliography}

\newpage 

\part*{Appendix}

\setcounter{section}{0}
\renewcommand{\thesection}{\Alph{section}}

\section{Tensor product construction}

\label{sec:app:tensor-pen}

We can build a tensor product basis as follows. If we have two univariate smooths $s_x(x)$ and $s_y(y)$, we can look at their basis expansions:
\begin{equation*}
    s_x(x) = \sum_k \beta_k b_k(x) \quad \text{and} \quad s_y(y) = \sum_l \zeta_l d_l(y),
\end{equation*}
where in each summation the Greek letters, $\beta_k$ and $\zeta_l$, indicate parameters to estimate and Latin letters, $b_k$ and $d_l$, give fixed basis functions. 

We can generate a bivariate function from a combination of these terms by taking $s_x(x)$ and asking how to make this a function of $x$ and $y$. We can do this by making each $\beta_k$ a function of $y$, using the basis expansion we already have for $s_y(y)$, hence:
\begin{equation*}
    \beta_k(y)  = \sum_l \zeta_{kl} d_l(y).
\end{equation*}
Then we can substitute that back into the basis expansion for $s_x(x)$ to give:
\begin{equation*} 
    s_{xy}(x,y) = \sum_k \beta_{k}(y) b_k(x) = \sum_k  \sum_l \zeta_{kl} d_l(y) b_k(x) .
\end{equation*}
We can iterate this process to build tensor products of any number of dimensions.

The corresponding penalties, which measure the wigglyness of the smooth, are constructed as follows. Primary reference for this is \cite[][Section 5.6.2]{wood_generalized_2017-1} and readers should consult that text for further details, including more general versions of the penalties.

Going back to the univariate terms we started with, each have a penalty:
\begin{equation}
    J_x(s_x) = \lambda_x \boldsymbol{\beta}^\intercal \mathbf{S}_x \boldsymbol{\beta} \quad \text{and} \quad J_y(s_y) = \lambda_y \boldsymbol{\zeta}^\intercal \mathbf{S}_y \boldsymbol{\zeta},
\label{eq:mat-pen}
\end{equation}
where $\boldsymbol{\beta}$ and $\boldsymbol{\zeta}$ are vectors containing the relevant parameters. $\mathbf{S}_x$ and $\mathbf{S}_y$ are fixed matrices measuring the wigglyness of the basis functions. To regulate the importance of the penalties, each has an associated \textit{smoothing parameter}, $\lambda_x$ and $\lambda_y$ which scale the influence of the penalty. $J_x(s_x)$ and $J_y(s_y)$ will be added to the negative $\log$-likelihood to create a penalized likelihood when taking a ``frequentist'' (actually empirical Bayes) view of the model, or may be viewed as precision matrices of the multivariate normal (zero mean) priors on the parameters \citep{millerBayesianViewsGeneralized}.

For the sake of clarity and simplicity, we'll say $J_x(s_x)$ and $J_y(s_y)$ are based on a cubic regression spline basis, so in each case we have:
\begin{equation}
    J_x(s_x) = \lambda_x \int \left( \frac{\partial^2 s_x(x)}{\partial x^2}\right)^2 dx \quad \text{and} \quad J_y(s_y) = \lambda_y \int \left( \frac{\partial^2 s_y(y)}{\partial y^2}\right)^2 dy.
\label{eq:int-pen}
\end{equation}
Intuitively, we can think of the second derivatives as calculating the how the smoother changes over the range of $x$, squaring this removes issues about the sign of those changes, and integrating that gives an overall wigglyness measure for the term. A bit of algebra shows that we can re-write (\ref{eq:int-pen}) in the form of (\ref{eq:mat-pen}) where we have fixed matrices ($\mathbf{S}_x$ and $\mathbf{S}_y$) we only need calculate once and parameters ($\boldsymbol{\beta}$ and $\boldsymbol{\zeta}$) which change during model fitting. In practice this means that model fitting has the computational complexity of matrix multiplication, rather than of differentiation/integration.

Now we want to find $J_{xy}(s_{xy})$, the combined penalty of the interaction between smooths of $x$ and $y$. We previously found that the basis for the tensor product of $s_x$ and $s_y$ is:
\begin{equation*}
    s_{xy}(x,y) = \sum_k  \sum_l \zeta_{kl} b_k(x) d_l(y).
\end{equation*}
A straightforward way of thinking about a penalty for this term (again thinking about the cubic spline case) is:
\begin{equation}
J_{xy}(s_{xy}) = \int \lambda_x \left( \frac{\partial^2 s_{xy}(x,y)}{\partial x^2}\right)^2 + \lambda_y \left( \frac{\partial^2 s_{xy}(x,y)}{\partial y^2}\right)^2 dx dy.
\label{eq:tensor-pen}
\end{equation}
This formulation lets us look at the changes in each direction, holding the other constant, so we get the overall wigglyness of the function from the sum over those directions.

We can re-write (\ref{eq:tensor-pen}) in the matrix form, as we did in (\ref{eq:mat-pen}). We can again iterate this process to build tensor products of any number of dimensions. \cite{wood_generalized_2017-1}, Section 5.6.2 provides further details.


\section{Distributed lag derivation}

\label{dist-lag-derivation}

We begin by defining the basis expansions for the independent smooths of $x$ and \texttt{lag}:
\begin{equation*}
    s_x(x) = \sum_{k=1}^K \beta_k b_k(x) \quad \text{and} \quad s_\texttt{lag}(\texttt{lag}) = \sum_{l=1}^L \zeta_l d_l(\texttt{lag}).
\end{equation*}
We would like to let coefficients $\beta_k$ depend on $\texttt{lag}$. We can use the tensor product trick we saw in Section \ref{sec:tensors}, by rewriting the coefficients for the smooth of $x$ in terms of the basis expansion for $\texttt{lag}$, as above:
\begin{equation*}
    \beta_k(\texttt{lag})  = \sum_{l=1}^L \zeta_{kl} d_l(\texttt{lag}).
\end{equation*}
Substituting that back into the equation for the smooth of $x$ (and adding $i$ subscripts for observations), giving:
\begin{equation*}
    s_{x,\texttt{lag}}(x_{i},\texttt{lag}_{i}) = \sum_{k=1}^K  \sum_{l=1}^L \zeta_{kl} b_k(x_{i}) d_l(\texttt{lag}_{i}).
\end{equation*}
Finally, we note that we will supply $x$ and \texttt{lag} as vector predictors, so we invoke the summation convention and sum over the lags (indexed by $t$) per observation to obtain
\begin{equation*}
    \sum_{t=1}^T s_{x,\texttt{lag}}(x_{it},\texttt{lag}_{it}) = \sum_{t=1}^T \sum_{k=1}^K  \sum_{l=1}^L \zeta_{kl} b_k(x_{it}) d_l(\texttt{lag}_{it}).
\end{equation*}

The technique used to characterize this simultaneous and interacting effect of the predictor time series intensities and indices is known in the distributed lag literature as the \textit{cross-basis}.

This class of models includes a lot of possibilities \citep{armstrongModelsRelationshipAmbient2006, pengBayesianHierarchicalDistributed2009, gasparriniDistributedLagLinear2011}, including various application-specific extensions. Here we only consider options which fit within the GAM framework provided by \mgcv{}. One advantage of this implementation is that we can use penalties to ensure that our models do not over fit (which are not always considered in the literature).

\end{document}